\documentstyle[prb,aps,preprint,tighten,epsfig]{revtex}
\draft
\begin{document}

\title {Effects of disorder on two strongly correlated coupled chains}

\author{E. Orignac and T. Giamarchi}
\address{Laboratoire de Physique des Solides, Universit{\'e} Paris--Sud,
                  B{\^a}t. 510, 91405 Orsay, France\cite{junk}}
                   \date{\today}
\maketitle

\begin{abstract}
We study the effects of disorder on a system of two coupled chain
of strongly correlated fermions (ladder system), using a renormalization
group technique. The stability of the phases of the pure system has been
investigated as a function of interactions both for fermions with spin
and spinless fermions. For spinless fermions the repulsive side is
strongly localized whereas the system with attractive interactions is
\emph{stable} with respect to disorder, at variance with the single
chain case. For fermions with spins, the repulsive side is also
localized, and in particular the d-wave superconducting phase found for
the pure system is totally destroyed by an arbitrarily small amount of
disorder. On the other hand the attractive side is again remarkably
stable with respect to localization. We have also computed the charge
stiffness, the localization length and the temperature dependence of the
conductivity for the various phases. In the range of parameter where
d-wave superconductivity would occur for the pure system the
conductivity is found to \emph{decrease} monotonically with temperature,
even at high temperature, and we discuss this surprising result.
For a model with one site repulsion and nearest neighbor attraction, the
most stable phase is an orbital antiferromagnet . Although this phase
has no divergent superconducting fluctuation it can have a divergent
conductivity at low temperature. Finally, to make comparison of our
results with experimental ladder systems, we treated the interladder
coupling in a mean field approximation. We argue based on our results
that the superconductivity observed in some of these compounds cannot be
a simple stabilization of the d-wave phase found for a pure single
ladder. Application of our results to systems such as quantum wires is
also discussed. In particular the
corrections to conductance in a two channel quantum wire have been obtained as a
function
of system length, temperature and interactions.
\end{abstract}
\narrowtext

\section{Introduction}

Strongly interacting systems constitute nowadays one of the most
challenging problem of condensed matter physics. In one dimension a
fairly complete solution of the interacting problem can be obtained, and
it is well known that  one dimensional systems are
one of the simplest realizations of non-Fermi liquids, and have generic
properties known as Luttinger liquids
\cite{haldane_bosons,haldane_bosonisation,solyom_revue_1d,emery_revue_1d}.
Prompted by a variety of experimental situations ranging from organic
conductors
to High Tc superconductors, there has been in the recent years,
a growing interest in systems of coupled
interacting electron chains. Unfortunately,
despite the good understanding of purely one
dimensional systems, the effects of interchain hopping, allowing to go
from one to higher (two or three) dimensions are much less known.
Whether non-Fermi liquid properties can be
retained even in presence of finite hopping or not is still a
highly controversial issue
\cite{clarke_coherence_coupled,boies_hopping_general}.

Many studies have therefore focused on systems of few coupled chains
\cite{dagotto_ladder_review}
(two coupled chains being the so-called ladder systems), for which much
more controlled analytical \cite{fabrizio_2ch_rg,kveschenko_spingap,finkelstein_2ch,%
schulz_2chains,balents_2ch,nagaosa_2ch,nersesyan_2ch,yoshioka_coupledchains_interaction}
or numerical \cite{dagotto_lanczos_2ch,noack_dmrg_2ch,poilblanc_2ch_mc,poilblanc_2ch,%
tsunegutsu_2ch} techniques can be applied
allowing for a deeper understanding of their physical properties.
For commensurate filling, i.e. one electron per site, the system becomes
equivalent to coupled spin chains, since the charge degrees of freedom
are frozen by a Mott transition. Important differences between ladders
with and even and odd number of legs were expected, in a way
reminiscent of the Haldane conjecture between one-dimensional systems
with integer and half integer spins. In particular
ladders with an even number of legs were predicted to have a spin gap.
Good experimental realizations of such coupled spin chains
like $\rm Sr_{n-1}Cu_{n+1}O_{2n}$ \cite{takano_srcuo,takano_spingap} and $\rm
VO_2P_2O_7$ \cite{eccleston_vo2p2o7,johnston_vo2p2o7} compounds have confirmed such
behavior.
Due to the presence of such a spin gap an even more spectacular effect is
expected upon doping. At the opposite of single chains, that exhibit
either a spin density wave or charge density wave ground state for
repulsive interactions, the ladder system  is believed to have a
superconducting ground state involving
pairing across the chains. That superconducting state has similarities
 with d-wave paring that
has been advocated in some two dimensional models of strongly
correlated electrons for High Tc
superconductors
\cite{scalapino_dwave,bickers_dwave1,bickers_dwave2,bickers_dwave3,bickers_dwave4,%
miyake_dwave,pines_hightc}
such as the existence of a spin gap and a sign change of the
superconducting order parameter when one moves on the ``Fermi
Surface''. In the strong coupling limit i.e. the t-J model, the d-wave
phase can also be viewed as a RVB state\cite{baskaran_jauge}.

However all the studies of ladder systems have been, up to now,
restricted to pure systems. Unfortunately (or maybe fortunately) it is
well known, that for one dimensional systems disorder has extremely
strong effects. For a non-interacting system, it is well known that
all states get localized in the presence of an infinitesimal random potential
\cite{abrikosov_rhyzkin,berezinskii_conductivity_log}. Interactions
can modify this picture, but for one chain system delocalization
occurs only for strongly attractive interactions.
In particular even normal s-wave superconducting phases are destroyed by
non magnetic impurities except for exceedingly attractive
interactions (see e.g. Ref.~\onlinecite{giamarchi_loc} and references therein), and
no Anderson's theorem exists even for
weakly coupled one-dimensional systems\cite{suzumura_mean_field}.

In order to compare the
theoretical predictions of d-wave superconductivity in doped ladder systems with
experiments, it is therefore of
prime importance to understand the effects of disorder on the phase
diagram of the pure ladder system. One of the important question is of
course the stability of the newly found d-wave superconducting phase
since there are no obvious reasons why it would survive the
introduction of a small amount of disorder.
Such a study is also
relevant to the physics of quantum wires with few
channels\cite{meirav_wires,meirav_gaas_1d,goni_gas1d,tarucha_wire_1d,kastner_coulomb}.
In quantum wire systems, the situation is however complicated by the
occurrence of long range Coulomb forces\cite{kastner_coulomb} that can induce a one
dimensional analog of the Wigner Crystal
\cite{glazman_single_impurity,schulz_wigner_1d},
which can drastically modify the response of the system to disorder
\cite{maurey_qwire}.
However, the presence of charges in the grids of the quantum wire
systems can be cleverly used
to screen completely the long range interactions, and have an
experimental realization of a Luttinger liquid \cite{tarucha_wire_1d}.
By changing the gate voltage it is possible to have more than one band at
the Fermi level, in a controlled way. The quantum wire is thus a possible
realization of a two (or more) leg ladder.
Interband tunneling plays the role of interchain hopping. They provide
ideal systems in which to check for the effect of disorder
\cite{goni_gas1d,tarucha_wire_1d}.

Besides the exciting possibility to test the ability of Luttinger liquid models to
describe accurately the now available
quasi one dimensional experimental systems, investigation
of disorder effects in ladders presents in its own right a great theoretical
interest. Indeed, the two chain
problem is the simplest one to study the effects of interchain hopping
onto the Anderson localization in presence of interactions, giving some
clues onto this difficult topic in more than one dimension.
In particular, one would be interested in obtaining boundaries between
localized and delocalized phases and the dependence on localization
lengths on disorder.
Another question of particular interest is the effect of interactions on
physical quantities controlled by disorder such as the conductivity for a
macroscopic system, or for a mesoscopic one the persistent
currents. In particular, for a one chain system, it was shown that for
a system with spin degrees
of freedom persistent current were enhanced
\cite{giamarchi_persistent_1d} by repulsive
interactions, at variance on what happened for a spinless system. It is
therefore important to check whether this striking results still holds in
a more two dimensional system.

In this paper, we consider the effects of a weak random potential scattering on
systems of coupled fermionic chains both with spin and spinless using bosonization
and RG techniques. A short account of some of the results of this paper were
presented in Ref.~\onlinecite{orignac_2chain_short}.
Besides giving the
phase boundaries, the RG also provides us with  expressions of
the localization lengths, temperature dependence of conductivity, and
dependence  of persistent currents with system size.
The plan of the paper is as follows~:

In section \ref{SPINLESS}, we  discuss the spinless fermions 2 legs ladder
problem . We first recall the phase diagram of the
 pure system \cite{nersesyan_2ch}, then  consider the effects of  disorder. This allows for a
detailed comparison of the transport properties of the 2 chain system
with the ones of the one chain spinless fermion system and the ones of
one chain of fermion with spin. We
show that contrarily to naive expectations, the ladder spinless fermions
system is very different from the one chain
system with spin, and that the effect of interactions on persistent
currents is even more violent on a two chain spinless fermions system
than it was for a one chain system.

In section \ref{2ch_Hubbard}, we  discuss the technically more
involved case of fermions with spins.
Following the same methodology, we first recall the phase diagram of
the non-random 2 chain system and then consider the effects of a weak
random potential on the phase diagram.
As for the
spinless problem, we give a detailed discussions of the transport
properties in the disordered phases. We compare these results with the
ones already known for one chain, and show that the reduction of
persistent current by attractive interactions that is observed in one
chain of fermions with spin should be almost absent in the two chain
system.  The d-wave superconductivity of
the  fermionic 2 chain system is a feature that is not preserved in the
presence of a very small amount of disorder. On the other hand for some
values of the parameter, and orbital antiferromagnetic phase exists.
This phase has an infinite conductivity  \emph{even in the
presence of disorder} although it has exponentially decaying
superconducting correlations.

In order to compare our results with
experiments on doped ladder systems, it is necessary to treat
interchain coupling which stabilizes superconductivity at finite
temperature in real systems, and may also reduce the sensitivity of the
system to disorder. Thus in section \ref{MEANFIELD}  we  examine the
mean field theory for
the d-wave superconductor in an array of coupled disordered ladders.
We give a criterion for persistence of
superconductivity in the presence of disorder and show that d-wave
superconductivity remains unstable except in very pure samples or in
the presence of a very strong Josephson coupling between ladders.

In section \ref{link_w_expts}, we summarize  the
implications for  experimental systems such as doped $SrCuO$ chains
and quantum wire with two channels. We
claim that in recently synthesized doped ladder systems, the physics
of the superconducting phase is more likely to be of two dimensional
origin rather than just a stabilization of a ladder d-wave
superconductivity. Conclusions can be found in
section~\ref{concl}. Finally most of the technicalities can be found in
the appendixes.

\section{Spinless fermions}\label{SPINLESS}

\subsection{Pure system}

\begin{figure}
\centerline{\epsfig{file=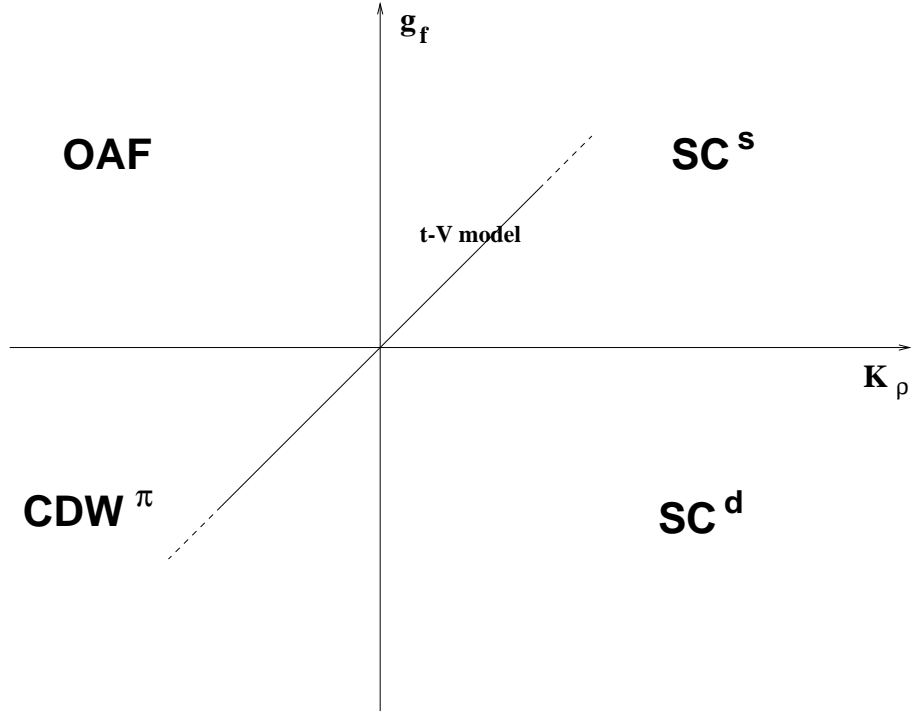,angle=-90,width=12cm}}
   \vspace{0.5cm}
\caption{The phase diagram of a generic spinless ladder in terms of
$g_f$ and $K_{\rho}$. $K_\rho > 1$ means attraction in the
symmetric charge sector and
$K_\rho < 1$ repulsion. The line depicts the phase spanned by the pure
t-V ladder, leading to a $CDW^\pi$ phase for $V>0$ and a superconducting
$SC^s$ phase for $V<0$.}
\label{pure_tV}
\end{figure}

Let us consider first two chains of spinless fermions
coupled by an interchain hopping $t_{\perp}$. For simplicity we first
consider only nearest neighbor interactions, and interchain
interactions. The effect of more complicated interactions will be
detailed below.
The Hamiltonian for the pure system reads
\begin{eqnarray}
\label{teve}
H & = & -t \sum_{i,p} c^{\dagger}_{i,p} c_{i+1,p} + \text{h.c.} +
V \sum_{i} n_{i,p} n_{i+1,p} \nonumber \\
 & & + t_{\perp} \sum_{i} c^{\dagger}_{i,1} c_{i,-1} + \text{h.c.}
       + U\sum_{i} n_{i,1} n_{i,-1}
\end{eqnarray}
where $p= -1,1$ is the chain index and $i$ is the site index. To treat
the interactions, it is convenient to rewrite the Hamiltonian in term of
boson operators \cite{emery_revue_1d,solyom_revue_1d,nijs_equivalence}.
To do so, we linearize the fermions dispersion relation
around $k_F$,
introduce right (R) and left movers (L) for each chain,
and take the continuum limit $c_{n,r,p}\rightarrow
\sqrt{\alpha}\psi_{r,p}(n\alpha)$ with $r=L,R $, $p=\pm1$
the chain index and $\alpha$ the lattice spacing.
We use the bonding $\psi_{o}=\frac{\psi_{1}+
\psi_{-1}}{\sqrt{2}}$
and anti-bonding $\psi_{\pi}=\frac{\psi_{1}-\psi{-1}}{\sqrt{2}}$
bands base and
introduce the densities $\rho_{r,o,\pi}(x)=:\psi_{r,o,\pi}^{\dagger}(x)
\psi_{r,o,\pi}(x):$. We then define the canonically conjugate
fields $\phi_{\rho,\parallel}$ and $\Pi_{\rho,\parallel}$ via
\begin{eqnarray} \label{fields}
\partial_{x}\phi_{\rho,\parallel} & = & -\frac{\pi}{\sqrt{2}}
(\rho_{L,o}+\rho_{R,o} \pm \rho_{L,\pi} \pm \rho_{R,\pi}) \\
\Pi_{\rho,\parallel} & = & \frac{1}{\sqrt{2}}(\rho_{R,o} \pm
\rho_{R,\pi}-\rho_{L,o} \mp\rho_{L,\pi})
\end{eqnarray}
and the field $\theta_{\rho,\parallel}(x)
=\int_{-\infty}^{x} \Pi_{\rho,\parallel}(x')dx'$.
More details on the
bosonization technique can be found in Appendix~\ref{bosonisation}.
In term of these fields the Hamiltonian becomes\cite{nersesyan_2ch}~:
\begin{eqnarray} \label{bosonise}
H & = & H_{\rho}+ H_{\parallel}, \qquad\qquad
H_{\rho} = \int \frac{dx}{2\pi}\left[ u_{\rho} K_{\rho}
(\pi \Pi_{\rho})^{2} +
\frac{u_{\rho}}{K_{\rho}}( \partial_{x} \phi_{\rho})^{2}\right]\\
H_{\parallel} &=& \int \frac{dx}{2 \pi} \left[u_{\parallel}
K_{\parallel} (\pi
\Pi_{\parallel})^{2}+\frac{u_{\parallel}}{K_{\parallel}}(\partial_{x}
\phi_{\parallel})^{2}\right] +\int dx t_{\perp} \frac{\sqrt{2}}{\pi}
\partial_{x}
\phi_{\parallel}\nonumber \\
 & &+ \int dx \left[\frac{2g_{\perp}}{(2 \pi \alpha)^{2}}
\cos(\sqrt{8}\phi_{\parallel})+ \frac{2g_{f}}{(2 \pi \alpha)^{2}}
\cos( \sqrt{8} \theta_{\parallel}) \right] \nonumber
\end{eqnarray}
For the microscopic Hamiltonian (\ref{teve}), one finds
\begin{eqnarray}
\label{K_ug}
K_\parallel=1+\frac{Ua}{2\pi v_F}\nonumber \\
u_\parallel=v_F(1-\frac{Ua}{2\pi v_F}+\frac{Va}{\pi v_F}(1-\cos(2k_F
a))) \nonumber  \\
g_f=-Va(1-\cos(2k_F a))\nonumber\\
g_\perp=Ua - Va(1-\cos(2k_F a))\nonumber\\
u_\rho = v_F(1+\frac{Ua}{2\pi v_F}+\frac{Va}{\pi v_F}(1-\cos(2k_F a)))
\nonumber  \\
K_\rho = 1-\frac{Ua}{2\pi v_F} -\frac{Va}{\pi v_F}(1-\cos(2k_F a))
\label{param}
\end{eqnarray}
with $v_F=2ta \sin(2k_Fa)$. Therefore
for the pure t-V model, one has $K_\rho<1$
(resp. $K_\rho>1$) and $g_f < 0$ (resp. $g_f > 0$) for repulsive (resp.
attractive) interactions and $K_{\parallel}=1$ for all $t,V$. In fact
(\ref{bosonise}) describes the most general two chain spinless system.
More complicated (i.e. longer range and interchain interactions) lead
only to a change in the parameters $K$, $u$ and $g$.
By adding interchain interactions such as $U$ in formula (\ref{param})
or longer range interactions one can in particular access the other
regimes $K_{\rho}>1$ and $g_{f}<0$ or $K_{\rho}<1$ and $g_{f}>0$.
The physics of the system is readily seen on (\ref{bosonise}).
The $t_{\perp}$ term suppresses $\cos(\sqrt{8}\phi_{\parallel})$.
Depending on the value of $K_{\parallel}$, the $\theta_{\parallel}$
can either remain massless or develop a gap. We concentrate here on
the case where
$\theta_{\parallel}$ develops a gap and
acquires a non-zero expectation value determined by minimizing the
ground state energy (see appendix~\ref{bosonisation}). This situation
always occur for the t-V model \cite{nersesyan_2ch}.
By mapping (\ref{bosonise}) on a problem of one
chain of fermions with spin and spin-anisotropic interactions in a
magnetic field \cite{giamarchi_spin_flop}, one can obtain the complete
phase diagram for the pure case \cite{nersesyan_2ch}. Since,
due to the one dimensional nature of the problem, no true ordered state
exists, one has to find the most divergent instability.
As for one chain, two main type of instabilities are possible:
particle-hole (density, current etc.) instabilities or particle-particle
(i.e. superconducting) ones. The operators with the most divergent
susceptibilities are in boson form
\begin{eqnarray}
O_{CDW^{\pi}}& = &
\psi^{\dagger}_{R,1}(x)\psi_{L,1}(x)-\psi^{\dagger}_{R,-1}
\psi_{L,-1}(x)\sim
e^{\imath\sqrt{2}\phi_{\rho}}\cos(\sqrt{2}\theta_{\parallel}),
\nonumber\\
O_{SC^s} & = &
\psi_{L,o}(x)\psi_{R,\pi}+\psi_{L,\pi}\psi_{R,o} \sim
e^{\imath\sqrt{2}
\theta_{\rho}}\sin(\sqrt{2}\theta_{\parallel}) \nonumber\\
O_{OAF} & = &
i(\psi^{\dagger}_{R,1}(x)\psi_{L,-1}(x)-\psi^{\dagger}_{R,-1}(x)
\psi_{L,1}(x))\sim
e^{\imath\sqrt{2}\phi_{\rho}}\sin(\sqrt{2}\theta_{\parallel}),
\nonumber\\
O_{SC^d}& = &
\psi_{L,o}\psi_{R,\pi}-\psi_{L,\pi}\psi_{R,o}
\sim e^{\imath\sqrt{2}\theta_{\rho}}\cos(\sqrt{2}\theta_{\parallel})
\nonumber
\end{eqnarray}
They describe respectively out of phase charge density waves,
an orbital antiferromagnetic phase and chain symmetric ``s'' and
antisymmetric ``d'' type superconductivity. The out of phase charge
density has a $2k_F$ modulation of the density along the chain and a
change of sign across the chains. In the
orbital antiferromagnet currents go from one chain
to the other with wavevector $2k_F$, giving  currents circulating
around plaquettes of length $\pi/k_F$. The superconducting phases are
the standard ones, given on the original model by
\begin{eqnarray}
O_{SC^d}(n)=c_{n,1}c_{n,2} \nonumber \\
O_{SC^s}(n)=c_{n+1,1}c_{n,1}-c_{n+1,2}c_{n,2} \nonumber \\
\end{eqnarray}
The most stable phase depends on the parameters $K$ and $g$. The various
cases are given in Table~\ref{table0}, and the phase diagram shown in
figure \ref{pure_tV}. In Ref.~\onlinecite{nersesyan_2ch} the bosonized forms
of $O_{SC^s}$ and $O_{SC^d}$ are exchanged due to the neglect of
anticommuting operators (see appendix~\ref{bosonisation}),
so that the two superconducting phases have been erroneously exchanged.

\subsection{Effects of disorder}

Now, we consider the effect of the disorder on
(\ref{teve}-\ref{bosonise}). We introduce
a random on-site potential $\epsilon_{i,p}$ uncorrelated from
site to site and from chain to chain:
\begin{equation}
\label{lattice_disorder}
H_{\text{random}}=\sum_{i \atop p=\pm 1} \epsilon_{i,p} c^\dagger_{i,p}c_{i,p}
\end{equation}
with $\overline{\epsilon_{i,p}\epsilon_{j,p'}}=D\delta_{i,j}\delta_{p,p'}$.
In the continuum limit and using the bonding antibonding basis the
disorder becomes
\begin{equation}
\label{cont_bands_disorder}
H_{\text{random}}=\int dx \left[\epsilon_s(x)(\psi^\dagger_0(x)
\psi_0(x)+\psi^\dagger_\pi(x)\psi_\pi(x)) +\epsilon_a(x)
(\psi^\dagger_0(x)\psi_\pi(x)+\psi^\dagger_\pi(x)\psi_0(x)) \right]
\end{equation}
with $\epsilon_{s,a}=(\epsilon_1 \pm \epsilon_{-1})/2$
and
$\overline{\epsilon_{\alpha}(x)\epsilon_{\beta}(x')}=\frac{D a}{2} \delta(x-x')
\delta_{\alpha,\beta}$. Using the expression of fermion operators
defined in Appendix~\ref{bosonisation} and (\ref{fields}) one obtains for
the disorder term
\begin{eqnarray}
\label {full_bosonized_random}
H_{\text{random}}  =  \int dx \left[\eta_s(x)\frac{\sqrt{2}}{\pi}\partial_x
\phi_\rho(x) +\frac{\xi_s(x)}{\pi\alpha}
e^{\imath \sqrt{2}\phi_\rho} \cos(\sqrt{2}\phi_\parallel) +
\frac{ \xi_s^*(x)}{\pi\alpha} e^{-\imath \sqrt{2}\phi_\rho}
\cos(\sqrt{2}\phi_\parallel)\right] & & \nonumber \\
 + \int dx \left[\frac{\eta_a(x)}{\pi\alpha} \cos(\sqrt{2}\phi_\parallel)
\cos(\sqrt{2}\theta_\parallel)+
\frac{ \xi_a(x)}{\pi\alpha} e^{\imath \sqrt{2}\phi_\rho}
\cos(\sqrt{2}\theta_\parallel)
+\frac{ \xi_s^*(x)}{\pi\alpha} e^{-\imath \sqrt{2}\phi_\rho}
\cos(\sqrt{2}\theta_\parallel)\right] & &
\end{eqnarray}
where the disorder has been split in a $q\sim 0$ component
($\eta_{s,a}$) and a $q\sim 2k_F$ one ($\xi_{s,a}$). As for one chain
the $\eta$ and $\xi$ are uncorrelated and
\begin{eqnarray}
\overline{\eta_{s,a}(x)\eta_{s,a}(x')} &=& D_{s,a}a \delta(x-x') \\
\overline{\xi_{s,a}(x)\xi_{s,a}(x')} &=& 0 \\
\overline{\xi_{s,a}(x)\xi_{s,a}^*(x')} &=& D_{s,a}a\delta(x-x')
\end{eqnarray}

The $q\sim 0$ (forward scattering) part of the disorder does not affect
the conductivity and cannot lead to localization
\cite{abrikosov_rhyzkin}, but could in principle
modify the phase diagram and in particular destroy the gaps of the pure
phase. As for one chain, one can eliminate the
$\eta_s\partial_x \phi_\rho$ by a transformation
$\phi_\rho \to \phi_\rho +\frac{\sqrt8 K_\rho}{u_\rho}\int dx \eta_s(x)$.
The only effect of this term is therefore to give an additional
exponential decay in the density-density correlation functions.

Due to the presence of a gap in $\theta_\parallel$ (see
table~\ref{table0}), the
$\eta_a(x)\cos(\sqrt{2}\phi_\parallel)\cos(\sqrt{2}\theta_\parallel)$
term is always suppressed at lowest order. It could however generate
relevant terms at higher order. However higher order terms are either
identical to backscattering terms already present in the Hamiltonian, or
adds random contributions to
$g_\perp \cos\sqrt8\phi_\parallel$ and $g_f \cos\sqrt8\theta_\parallel$.
At small disorder these contributions are negligeable, and one can
completely disregard the forward scattering. We can therefore keep only
for the coupling to disorder $H_s+H_a$
\begin{eqnarray}
H_{s} & = & \int\frac{dx}{\pi\alpha}\xi_{s}(x)
e^{\imath\sqrt{2}\phi_{\rho}}
\cos(\sqrt{2}\phi_{\parallel}) + \text{h.c.} \label{symmetric} \\
H_{a} & =& \int\frac{dx}{\pi\alpha}\xi_{a}(x)
e^{\imath\sqrt{2}\phi_{\rho}}
\cos(\sqrt{2}\theta_{\parallel}) + \text{h.c.} \label{antisymm}
\end{eqnarray}
In $H_s$ the symmetric part of the disorder couples to the
in-phase charge density wave order parameter
$O_{CDW^0}=\frac{e^{\imath\sqrt{2}\phi_\rho}}
{\pi\alpha}\cos(\sqrt{2}\phi_\parallel)$, whereas the antisymmetric
part involves
$O_{CDW^\pi}$. Due to the gap in $\theta_{\parallel}$,
$\phi_{\parallel}$ has huge quantum fluctuations,
and consequently the symmetric part of the disorder $D_{s}$ is always
less relevant than the antisymmetric one $D_{a}$.
We can therefore focus on the latter and forget about the former.
The effect of (\ref{antisymm}) again depends on the values of $g_f$ and
$K$.

\subsubsection{$g_{f}<0$}

For $g_{f}<0$ (i.e. $V>0$ for the t-V model)
we can replace $\cos(\sqrt{2}\theta_{\parallel})$ by its
(non-zero) mean value and the coupling to disorder (\ref{antisymm})
reduces to $C \int dx
\xi_{a}(x)e^{i\sqrt{2}\phi_{\rho}(x)}+ \text{h.c.}$, where $C$ is a
constant. The effect of such a term can be determined, as for a single
chain\cite{giamarchi_loc}, by using a renormalization group (RG) procedure.
Upon varying a cutoff $\alpha$, similar to a lattice spacing in the
original lattice problem, one find the following renormalization for the
disorder
\begin{eqnarray}
\frac{d K_\rho}{d l} &=& - C_2 D_a \\
\frac{d D_a}{d l} &=& D_a (3 - K_\rho) \label{growth}
\end{eqnarray}
where $l=\ln(\alpha)$ and $C_2$ a constant. (\ref{growth}) implies a
localization-delocalization transition at $K_{\rho}=3$. For $K_{\rho}>3$
the disorder is irrelevant and the corresponding phase in the pure system
is stable. For $K_{\rho} < 3$ disorder grows. Although the system flows
to a strong coupling fixed point, it is natural \cite{giamarchi_loc}
to interpret this phase as localized by disorder,
since the disorder will pin the massless field $\phi_\rho$.
As a consequence, the d-wave superconducting phase is unstable
in the presence of disorder except for huge attractive interactions.
In the case of the t-V model at $V>0$, we have $K_{\rho}<1$ and
therefore the $CDW^{\pi}$ is always pinned by the disorder.

Similarly to the one chain problem the localization length can be
computed using the RG. For very weak disorder and far from the
transition one can neglect the renormalization of the exponent $K_{\rho}$
induced by $D_a$.
Using that approximation, we obtain:
\begin{equation}
D_a(l)=e^{(3-K_\rho)l}D_a(0)
\end{equation}
For $D_a(l)\sim v_F^2/\alpha$ that scheme breaks down and we have a strongly
disordered system. For such a system the localization length, i.e. the
scale of variation of the phase $\phi_\rho$ is of the order of the
(renormalized) lattice spacing $\alpha^*$. This occurs for
$e^{l^*}\sim
\left(\frac{v_F^2}{D_a(0)\alpha}\right)^{\frac{1}{3-K_\rho}}$.
Therefore
\begin{equation}\label{loc_length_gf<0}
L_{\text{2 ch.}}=\alpha(0)\left(\frac{v_F^2}{D\alpha}\right)^{\frac{1}{3-K_{\rho}}}
\end{equation}
Let us recall that for a non-interacting system, the localization
length is of the order of the mean free path
i.~e. $L_{\text{loc.}}\sim \frac{v_F^2}{ D}$.

Using the renormalization equation it is also possible \cite{giamarchi_loc}
to obtain the temperature dependence of the conductivity for
temperatures above the the pinning temperature $u/L_{\text{2 ch.}}$.
Below the pinning temperature, the conductivity is expected to decrease as
$\exp -\left(T_{\text{pin.}}/T\right)^\mu$, by analogy with non-interacting electrons.
A derivation of the temperature dependence of conductivity has been given
in Ref.~\onlinecite{giamarchi_loc}. Another method to derive the
temperature (or frequency) dependence of the conductivity is given in
appendix~\ref{condumem}. If one neglects the
renormalization of the exponents the conductivity behaves as
\begin{equation}
\sigma(T) \sim T^{2-K_\rho}
\end{equation}
Therefore, for $K_\rho <2$, the conductivity decreases, and there
is {\bf no remnant} of any superconducting behavior effect well above the
temperature at which the system is effectively pinned $T_{\rm pin} \sim
u/L_{\text{2 ch.}}$. Thus
the existence of  d-wave superconductivity in the pure system affects
the transport properties of the disordered system only for quite large
attraction. Analogous effects will occur for fermions with spins as will
be discussed in section ~\ref{2ch_Hubbard}.

\subsubsection{$g_{f}>0$}

For $g_{f}>0$ (i.e. attractive
interactions for a t-V model),
$\langle \theta_{\parallel} \rangle = \frac{\pi}{\sqrt{8}}$ and in
a first approximation the coupling (\ref{antisymm}) vanishes.
Obviously, this approximation is too crude and one
must integrate the fluctuations of $\theta_{\parallel}$ around its
mean value to get the effective coupling. This is done in
appendix~\ref{gapgap} and gives the following effective action
for $\phi_{\rho}$:
\begin{eqnarray}
\label{action-finale}
S_{\rho} &=& \int dx d\tau\left[\frac{(\nabla\phi_{\rho})^{2}}{2\pi K_{\rho}}+
(\xi(x)e^{\imath\sqrt{8}\phi_{\rho}(x,\tau)}+ \text{h.c.}) \right]
\end{eqnarray}
with $\overline{\xi(x)\xi^{*}(x')}=D\delta(x-x')$
and $D\sim D_{a}^2$.

The renormalization of the disorder is  given by an equation
similar to (\ref{growth}):
\begin{equation}
\frac{dD}{dl}=(3-4K_\rho)D(l)
\end{equation}
The disorder is now relevant only for $K_{\rho}<3/4$, leading
to three different phases for $g_{f}>0$: a random orbital
antiferromagnet for $K_{\rho}<3/4$, an ordered orbital antiferromagnet for
$3/4<K_{\rho}<1$ and a s-wave superconducting phase for $K_{\rho}>1$.
For the t-V model, $K_\rho > 1$, and the
``s''-wave superconducting phase is therefore {\em stable} with respect
to weak disorder, at variance to the single chain problem.
For the latter the delocalization only occured for {\em extremely}
attractive interactions i.e. $K_\rho > 3/2$. For the two chains problem
the localization-delocalization transition
arises in the immediate vicinity of the non-interacting point.
Contrarily to the case of repulsive interactions, interchain hopping
now strongly reduces
the localization effects.

The localization length in the random orbital antiferromagnet,
is now given by
\begin{equation} \label{loc_length_gf>0}
\frac{L_{\text{2 ch.}}}{\alpha}=(1/D)^{\frac{1}{3-4K_{\rho}}} =
\left(\frac{v_F^2}{D_a\alpha}\right)^{\frac{2}{3-4K_{\rho}}}
\end{equation}
The conductivity behaves both in the OAF and the s-wave phase as
\begin{equation}
\sigma(T) \sim T^{2-4 K_\rho}
\end{equation}
diverges as $T\to 0$ since the ground state is superconducting.
It is to be noted that although the OAF has no superconducting order
parameter, its conductivity can also be divergent for $K_\rho > 3/4$
{\em even} in the presence of disorder. An expanded discussion of
Orbital Antiferromagnet phases can be found in \ref{conductivity_2ch_w_spin} and
appendix \ref{other_perturbations}.

The resulting phase diagram is summarized on figure~\ref{disordered_tV},
together with the single chain phase diagram.

\subsection{Physical consequences}

\begin{figure}
\centerline{\epsfig{file=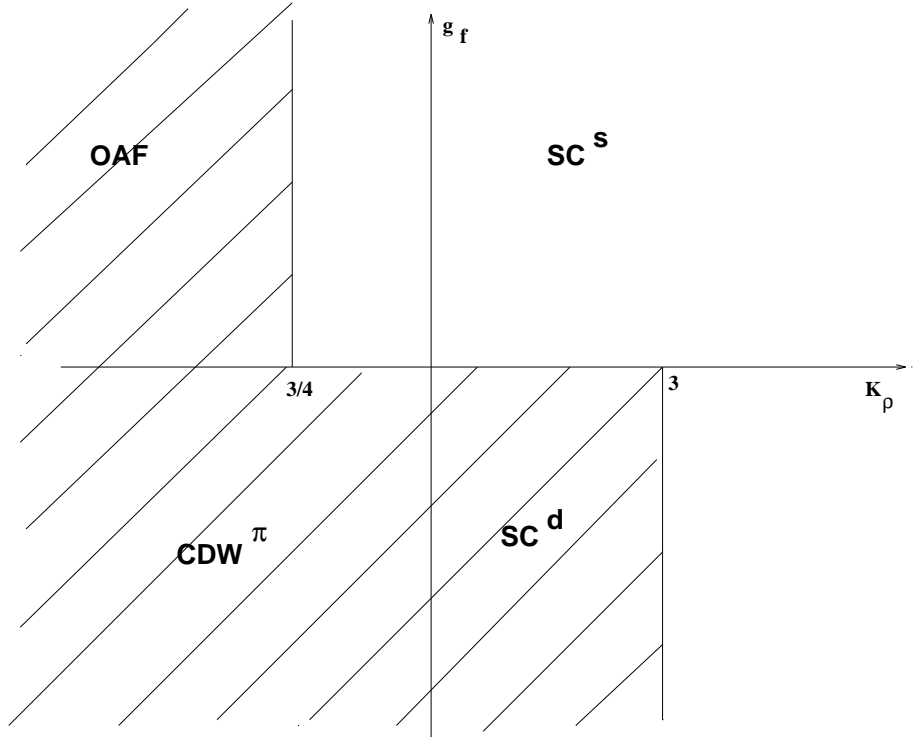,angle=-90,width=12cm}}
   \vspace{0.5cm}
\caption{The phase diagram of the disordered 2 chain t-V model in
terms of $g_f$ and $K_{\rho}$. For a single chain the system is
localized for $K_\rho < 3/2$. Ladder effect therefore {\em delocalize}
for attractive interactions and {\em enhance} localization for
repulsive ones.}
\label{disordered_tV}
\end{figure}

The ladder system shows drastically different sensitivity to disorder
depending on the sign of $g_f$: at $g_f<0$ localization effects are much
stronger than at $g_f>0$. This is obvious both on the phase diagram
shown on figure~\ref{disordered_tV}, and on the expression
(\ref{loc_length_gf<0}) and (\ref{loc_length_gf>0}) for the localization
length. For the case of a pure t-V model, $g_f>0$ $K_\rho>1$
when $V<0$ (attractive interactions) and as can be seen from
figure~\ref{disordered_tV} the system is delocalized. Although our
calculation do not allow us to come arbitrarily close to the
$V=0$ point for finite disorder, since the disorder has to be smaller
than the gaps of the pure system, we see that if we have a very small
disorder, the insulator superconductor
transition does occur in the vicinity of the non-interacting point.
This is remarkable and in marked contrast
with the single chain system where the delocalization
transition occurs for $K=3/2$ i.e. very strongly attractive
interactions even for arbitrarily weak disorder.
One could naively think that this effect is simply a manifestation of
the delocalization effect seen for non-interacting electrons when one
increases the number of channel (or the number of chains).
The mechanism is more
subtle however, and in in fact controlled by the interactions. Contrarily
to the noninteracting case where the localization length is simply
proportional to the number of chains, we have here a {\bf complete}
delocalization of the attractive region, and the localization length
becomes infinite.

For the repulsive case $V > 0$ (i.e. $g_f<0$ $K_\rho<1$) the opposite
effect occurs and
ladder system is {\em more} localized than the corresponding one chain
system. Indeed for one chain the localization length is given by
\cite{giamarchi_loc,suzumura_scha}
\begin{equation}
\frac {L_{\text{1 ch.}}} \alpha \sim \left(\frac{v_F^2}{D\alpha}
\right)^{\frac{1}{3-2K}}
\end{equation}
and is therefore longer than the one of the ladder system shown in
(\ref{loc_length_gf<0}). For very large repulsion ($K\to 0$) these two
length give back the standard Fukuyama-Lee pinning length of classical
charge density waves\cite{fukuyama_pinning}.
For finite repulsion the localization length of
the ladder system is much shorter than the one of the corresponding
one-dimensional system with the same $K$. Close to the non interacting
point $K\sim 1$, the localization length of open chain is just the mean
free path $L_{\text{1 ch.}} \sim v_F^2/D$ whereas the ladder one is
$L_{\text{2 ch.}}\sim \alpha \sqrt{\frac{v_F^2}{D \alpha}}$.

This peculiar behavior of the spinless
ladder system is due to the
gaping of some charge modes, that is different depending on whether the
interaction is attractive or repulsive. For the repulsive side $2 k_F$
charge fluctuations are still there and the gap
just reduces some of the quantum fluctuation and hence reinforce the
effects of disorder, whereas for the attractive side the gap kills the
dominant charge fluctuation coupled to disorder and helps to delocalize.
The sensitivity to disorder is therefore {\bf not} directly related to
the presence or absence of superconducting fluctuations in the pure
system, but more on how the {\bf density} fluctuations behave. The
smoother are the density fluctuations, the less localized the system is.
These effects will be even more transparent for the system with spins as will
be examined in details in section~\ref{2ch_Hubbard}.
As a consequence the transport properties {\em cannot} simply be guessed
by looking at the phase diagram of the pure system. They even can be
opposite to what our intuition based on higher dimensional system could
suggest: the more ``superconducting'' the system is the better is the
transport (see e.g. section~\ref{transport}).

\subsection{Persistent currents in the ladder
system}\label{persistent_spinless}

In addition to the temperature dependence of the conductivity,
one can compute the charge stiffness of the system
\cite{kohn_stiffness,shastry_twist_1d,scalapino_stiffness}
$D$, which
measures the strength of the Drude peak in a macroscopic system
$\sigma(\omega) ={\mathcal D} \delta(\omega) + \sigma_{\rm reg}$.
The stiffness ${\mathcal D}$ can be related to the change of the energy of the
ground state of the system in presence of an external flux by
\begin{equation}
{\mathcal D} = \frac{L}2 \left. \frac{d^2 E_0}{d \phi^2} \right|_{\phi=0}
\end{equation}
$E_0$ being the ground state energy of a ring in a field. $\phi$
denotes the boundary angle $\phi = 2 \pi f / f_0$ where $f$ is the flux
threading the ring and $f_0 = h c/e$ is the flux quantum.  This quantity
is directly related to the persistent currents for a mesoscopic system
\cite{buttiker_permanent_current,%
cheung_noninteracting_persistent,trivedi_conductivite_kubo,%
bouchiat_noninteracting_persistent,altshuler_noninteracting_persistent}
For a mesoscopic
system, the persistent current measures the response to a
finite flux by
\begin{equation}
J = L \left. \frac{d E_0}{d \phi} \right|_{\phi}
\end{equation}
Therefore the stiffness ${\mathcal D}$ provides a measure of the
persistent
currents for small (or close to a multiple of $2\pi$) flux since
$J =  2{\mathcal D} \phi$. Although the complete calculation of the
persistent currents at finite flux is also possible for a one
dimensional
interacting system, the calculation is more complicated in the presence
of disorder, and the stiffness carries enough information for our
present purposes.

The effects of interactions on persistent currents is an extremely
difficult question to answer in two or three dimensions.
Perturbative calculations suggests that interactions could enhance
persistent currents \cite{ambegaokar_interactions_current,%
muller-groeling_spinless_current,muller-groeling_spinless_2d,%
ramin_perturbative}. For a single spinless chain the persistent currents
were found to {\bf decrease} with more repulsive interactions
\cite{bouzerar_spinless_currents,bouzerar_rg_current,%
berkovits_coulomb_current}. This effect can naturally be explained using
a renormalization group technique, and it
 was shown that such
behavior is peculiar to the spinless problem and that for a single
chain of electrons with spins persistent currents should be enhanced by
repulsive interactions \cite{giamarchi_persistent_1d,bouzerar_rg_current}.
For the ladder system it is therefore very interesting to see if the
same effects occur and in particular to check again for the differences
between the spinless system and the system with spins. In particular one
could imagine that the chain index acts in a similar way than a spin
index for a single chain. As we will see this idea is far too naive. We
examine the spinless system in this chapter and the system with spins
will be investigated in chapter~\ref{2ch_Hubbard}.

For the ladder system, the conductivity stiffness
\cite{kohn_stiffness}is obtained using (\ref{stiffness_bos}) as
${\mathcal D}=2u_\rho K_\rho$. The factor of two
compared to the single chain expression (\ref{stiffness_bos}) is due
to the fact that
there are twice as much degrees of freedom in the 2 chain system. In the
following, we consider a finite
system, the size L  of which is smaller than the localization length.

 From the renormalization group equation for $u_\rho,K_\rho$
\cite{giamarchi_loc}, one can obtain \cite{giamarchi_persistent_1d} the
renormalization group equation for ${\mathcal D}$
\begin{equation}
\label{renormalise_stiffness}
\frac{d{\mathcal D}}{dl}=-D(l)
\end{equation}
  The conductivity stiffness of a disordered system of size L,
${\mathcal D} (L)$ is then
obtained by stopping the RG equation at $\alpha(l)=L$ and taking
${\mathcal D} (L)={\mathcal D} (l)$.
In the case $g_f<0$, we have seen that $D(l)=D(0)e^{(3-K_\rho)l}$,
at least
when $\alpha(l) \ll L_{\text{2 ch.}}$. Putting that approximation for $D(l)$
in (\ref{renormalise_stiffness}) gives us:
\begin{equation}
\label{approx_stiff_repuls}
{\mathcal D} (L)={\mathcal D} (0)-{\mathcal C}D(0)
\left[\left(\frac L {\alpha(0)}\right)^{3-K_\rho}-1\right]
\end{equation}
Using the expression for $L_{2ch.}$, (\ref{approx_stiff_repuls})
simplifies for length smaller than the localization length into
into
\begin{eqnarray}
{\mathcal D}_{g_f<0}(L) &=& {\mathcal
D}(0)-{\mathcal C}\left[\left(\frac L {L_{2 ch.(g_f<0)}}
\right)^{3-K_\rho}-1\right] \\
{\mathcal D}_{g_f>0}(L) &=& {\mathcal
D}(0)-{\mathcal C'}\left[\left(\frac L {L_{2 ch.(g_f>0)}}
\right)^{3-4K_\rho}-1\right]
\end{eqnarray}
Thus for $g_f>0$ the reduction of the stiffness is
less important than for $g_f<0$.

Therefore, the length dependence of the conductivity stiffness (and the
persistent currents) is extremely sensitive to the attractive or
repulsive
character of the interactions for the t-V model or any model with
intrachain-only interactions.
By comparison with the one chain case
\cite{giamarchi_persistent_1d,bouzerar_spinless_currents},
we see that
the effects of the interactions on the conductivity stiffness are
qualitatively the
same (i. e. repulsive interactions help in reducing the conductivity stiffness,
while attractive interactions reduce the decrease of conductivity stiffness by
disorder) but they are much stronger for two chains than for one chain.
In fact, for a t-V model, the reduction of conductivity stiffness would be
{\bf finite} for attractive interactions, even in an infinite system
since then the disorder is completely irrelevant.

It is noteworthy that the chain index does {\bf not} act in a similar
way as a spin degree of freedom, for which there would be an {\bf
increase} of the persistent currents showing again the important
difference between a system with and without spin
\cite{giamarchi_persistent_1d}. The physical reasons for this difference
are examined in more details in the next section.

\subsection{Spinless ladder vs. one chain with spin}

Naively, one could think that going from one chain to two chain amounts
to having one internal degree of freedom that is equivalent to spin,
and thus that the results for the system with spin will apply
straightforwardly to the ladder system. However, from what we have
seen precedingly, this is definitely not the case. In fact, we have
properties for the spinless ladder that are just the
contrary of the ones of the fermions with spin. Attractive
interactions delocalize in the spinless fermions case, whereas they
increase localization in the case of fermions with spin.
Persistent currents are enhanced for more attractive interactions in the
spinless ladder whereas repulsive interactions would enhance the
persistent currents\cite{giamarchi_persistent_1d} in a spin system.
The reason
for that is that the spinless ladder has no SU(2) symmetry
(except for $V=0$)
contrarily to one chain with spin. The minimum of the
ground state energy of the spinless ladder
corresponds to states that break the SU(2) symmetry because
$t_\perp$ plays the role of
a magnetic field \cite{nersesyan_2ch,giamarchi_spin_flop}. Thus such
phases cannot be obtained in an isotropic system of fermions with spin.

For attractive interactions, the only way for the symmetric
fermions with spin system to preserve SU(2) symmetry is to form
singlet phases such as $2k_F$ charge density waves or singlet superconducting
state. Coupling the charge density wave fluctuations with a random
potential implies strong localization effects. On the other hand, the
spinless ladder simply form pairs along the chains and
can avoid to form $2k_F$ fluctuations. Translated in the spin language, such
phase would be an anisotropic triplet superconductor with a spin  gap,
and would be forbidden by symmetry.
In the same way, for repulsive interactions, preserving SU(2) symmetry
prevents the formation of a gap, whereas a gap formation is possible for
the spinless ladder giving an out of phase charge density wave. In the spin
language, this corresponds to an anisotropic SDW.

Adding random potentials to the spinless ladder
results in a rather artificial model of fermions in a random potential
and a random field parallel to the z axis. Because of the anisotropy,
the system is more sensitive to the random field parallel to the z
axis than to the random potential.
Thus, for repulsive interactions, the anisotropic system has a very
strong coupling to disorder, whereas for repulsive interactions, it is
only weakly coupled. On the other hand, the isotropic system is only
feeling a random potential. When interactions are attractive, there is
a spin gap and CDW fluctuations that can couple to disorder, making
the system more localized. When interactions are repulsive, on the
other hand, there is no spin gap thus reducing the coupling of the CDW
fluctuations with disorder.

We conclude that for interacting systems, contrarily to their
non-interacting counterparts, not only the number of available
internal degrees of freedom but also the internal symmetries determine
the response to
random perturbations. Loosing some symmetries allows for a larger
variety of ground states, and thus to very different responses to weak
perturbations.

\section{Fermions with spin}
\label{2ch_Hubbard}

\subsection{Pure system}

The pure case has been analyzed in great details both analytically
\cite{fabrizio_2ch_rg,kveschenko_spingap,finkelstein_2ch,%
schulz_2chains,balents_2ch,nagaosa_2ch}
and numerically \cite{dagotto_lanczos_2ch,noack_dmrg_2ch,poilblanc_2ch,%
tsunegutsu_2ch}.
A very interesting feature of that model is the existence of a ``d-wave''
superconducting phase for purely repulsive interactions and the existence of a
spin gap.
The Hamiltonian is in the extended Hubbard case:
\begin{eqnarray}
H & = & -t\sum_{i,\sigma,p}c^{\dagger}_{i+1,\sigma,p}c_{i,\sigma,p} + \text{H. c.}
 - t_{\perp}\sum_{i,\sigma,p}c^{\dagger}_{i,\sigma,p}c_{i,\sigma,-p}
 \nonumber \\
 & & +U\sum_{i,p} n_{i,\uparrow,p}n_{i,\downarrow,p} +V \sum_{i,p} n_{i,p}n_{i+1,p}
\end{eqnarray}
with $p=\pm 1$ is  the chain index and $\sigma= \uparrow,\downarrow$
labels the spin.
In order to treat this Hamiltonian using bosonization
one has to separate the bonding $o$ and antibonding $\pi$ bands as was done
for spinless fermions.
Then, within each band, one can apply the standard  bosonization
formulas for fermions with spins.
As a consequence, the system is described by 4 fields $\phi_{\rho}^{\pi},
\phi_{\sigma}^{\pi}\phi_{\rho}^{o},
\phi_{\sigma}^{o}$ instead of 2 in the spinning case.
For the pure case we follow closely the derivation of
Ref.~\onlinecite{schulz_2chains}.
It is convenient in the following to replace the fields $\phi_{\nu}^{o,\pi}$
($\nu=\rho,\sigma$) by linear combinations:
$\phi_{\nu\pm}=\frac{1}{\sqrt{2}}(\phi_{\nu,o}\pm \phi_{\nu,\pi})$
The low energy physics depends on the signs of two
constants $g_1 , g_2$.
Physically, $g_2$ represents the forward scattering interaction, while
$g_1$ represents the backward scattering interactions.
The Hamiltonian consists of a free part~:
\begin{equation}
\label{free_2chains_Hubbard}
H=\sum_{\nu=\rho,\sigma \atop r=\pm}\int \frac{dx}{2\pi}\left[u_{\nu
r}K_{\nu r}
(\pi \Pi_{\nu r})^2 +\frac{u_{\nu r}}{K_{\nu r}}(\partial_x \phi_{\nu
r})^2 \right]
\end{equation}
and two sine-Gordon like part, one associated with interband processes
induced by
intrachain forward scattering~:
\begin{equation}
\label{eq:h2}
H_{int,2} = \frac{g_2}{2(\pi\alpha)^2} \int dx \cos2 \theta_{\rho-}
(\cos 2\phi_{\sigma-}+\cos 2 \theta_{\sigma-})
\end{equation}
The other associated with the intrachain backward scattering~:
\begin{equation}
\label{eq:back}
 H_{int,1}  =  \frac{2g_1^*}{(2\pi\alpha)^2}\int dx \left[\cos2\phi_{\sigma+}
(\cos 2\theta_{\rho-} + \cos 2\phi_{\sigma-} +\cos 2\theta_{\sigma-}) -
\cos 2\theta_{\rho-} \cos 2\theta_{\sigma-} \right]
 \end{equation}
In all cases, only one
of the four bosonic fields ($\phi_{\rho +})$ is gapless
\cite{schulz_2chains} and all physical quantities
depend on a parameter $K_{\rho+}$ of the symmetric charge mode,
analogous to the $K_\rho$ of the spinless problem.
In terms of $g_1,g_2$,$ K_{\rho+}$ is given by~:
\begin{equation}
K_{\rho+}=\left( \frac{2\pi v_F+ (g_1-2g_2)}{2\pi v_F- (g_1-2g_2)}\right)^{1/2}
\end{equation}
That expression is valid for the generic g-ological model. For the
extended Hubbard
model, we can go further as $g_1,g_2$ can be expressed in terms of
$U,V,k_F$  as~:
\begin{eqnarray} \label{g1g2=fUV}
g_1=Ua+2Va\cos(2k_Fa)\nonumber \\
g_1-2g_2=-(Ua+2Va(2-\cos(2k_Fa))
\end{eqnarray}
where $a$ is the lattice spacing.
The mean values of the three other fields are determined by
minimizing the energy of the ground state. Depending on the interactions
one can distinguish four sectors that are summarized in table
\ref{table1}

\begin{figure}
\centerline{\epsfig{file=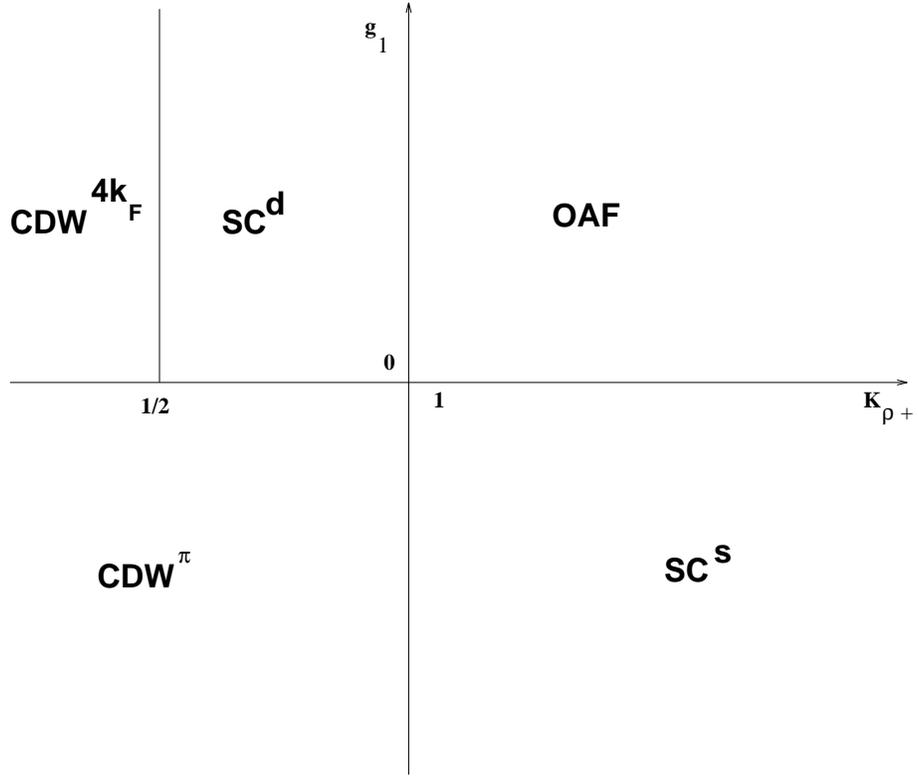,angle=-90,width=12cm}}
   \vspace{0.5cm}
\caption{The phase diagram of the pure 2 chain Hubbard model in terms
of $g_1$ and $K_{\rho+}$. $K_{\rho+}>1$ and $g_1<0$ corresponds to
purely attractive interactions. $K_{\rho+}<1$ and $g_1>1$ corresponds
to purely attractive interactions. For a Hubbard model, this leads to
a $SC^d$ phase for $U>0$ and a $SC^s$ phase for $U<0$.
}
\label{pure_hubbard}
\end{figure}

As for the spinless case one has to consider the various operators with
divergent susceptibilities
\begin{eqnarray}
O_{CDW^\pi}(n) =  \sum_{p,\sigma}p c^\dagger_{n,\sigma,p}c_{n,\sigma,p}
\label{lattice_cdwpi} \\
O_{OAF}(n) = \sum_{p,\sigma} p c^\dagger_{n,\sigma,p}c_{n,\sigma,-p}
\label{lattice_OAF} \\
O_{SC^s}(n) =  \sum_{p} c_{n,\sigma,p}c_{n,-\sigma,p} \label{lattice_scs} \\
O_{SC^d}(n) =  \sum_p c_{n,\sigma,p}c_{n,-\sigma,-p} \label{lattice_scd}
\end{eqnarray}
When taking the continuum limit these expressions become
\begin{eqnarray}
O_{CDW^\pi}=\sum_\sigma (\psi_{L1\sigma}^\dagger \psi_{R1\sigma} -
\psi_{L-1\sigma}^\dagger
\psi_{R-1\sigma})     \\
O_{OAF}= \imath\sum_\sigma (\psi_{L1\sigma}^\dagger \psi_{R-1\sigma} -
\psi_{L-1\sigma}^\dagger
\psi_{R1\sigma}) \\
O_{SC^s}=\sum_\sigma \left( \psi_{L0\sigma}\psi_{R0,-\sigma} +
\psi_{L\pi\sigma}\psi_{R\pi,-\sigma}\right) \\
O_{SC^d}=\sum_\sigma \left( \psi_{L0\sigma}\psi_{R0,-\sigma} -
\psi_{L\pi\sigma}\psi_{R\pi,-\sigma}\right)
\end{eqnarray}
where for the SC operators, one has to retain the $q\sim 0$ component,
whilst for the OAF and $CDW^\pi$  the $q\sim 2k_F$
component gives the dominant contribution.
To get the correct bosonized expression one has to pay extra care to the
anticommuting $U$ operators \cite{schulz_unpublished} and one obtains
\begin{eqnarray}
O_{CDW^\pi}=\frac{2}{\pi\alpha} e^{\imath\phi_{\rho+}}
\cos\phi_{\sigma+}\sin\theta_{\sigma-}    \\
O_{OAF}= \frac{2\imath}{\pi\alpha} e^{\imath\phi_{\rho+}}
\sin\phi_{\sigma+}\cos\theta_{\sigma-}\\
O_{SC^s}= \frac{2}{\pi\alpha}e^{-\imath\theta_{\rho-}}
\cos\phi_{\sigma+}\cos\phi_{\sigma-}\\
O_{SC^d}= \frac{2}{\pi\alpha}e^{-\imath\theta{\rho+}}
\sin\phi_{\sigma+}\sin\phi_{\sigma-}
\end{eqnarray}
 From the bosonized form of these operators (simplified by the fact that
$\langle \theta_{\rho-}\rangle=0$) everywhere and the expressions
given in table~\ref{table1} one can deduce
that sector I is a $SC^d$ phase, sector II an $OAF$ phase, sector III a
$SC^s$ phase and sector IV a $CDW^\pi$ phase.  The phase diagram of the
pure system is summarized in figure \ref{pure_hubbard}.
Note that for the pure Hubbard model, which corresponds to $V=0$ in
(\ref{g1g2=fUV}), one can only have the $SC^d$ phase (for $U>0$) or
the $SC^s$ phase (for $U<0$). The other phases could be obtained for
a more general model such as the extended Hubbard model. We will come
back to that point later.

\subsection{Effects of disorder}

Let us now add a weak random on-site potential~:
\begin{equation}
H_{\text{random potential}}=\sum_{i,\sigma,p}\epsilon_{i,p}n_{i,\sigma,p}
\end{equation}
with
$n_{i,p}=c^\dagger_{i,\uparrow}c_{i,\uparrow}+c^\dagger_{i,\downarrow}
c_{i,\downarrow}$
and $\overline{\epsilon_{i,p}\epsilon_{j,p'}}=D \delta_{i,j} \delta_{p,
p'}$.
We go through the same steps as in the spinless fermions section.
We got to the continuum limit, introduce the bonding and antibonding band,
and bosonize the resulting coupling to disorder.
Let us first consider the $q\sim 0$ part of the coupling to disorder.
For the symmetric part of the disorder this coupling is of the form~:
\begin{equation}\label{random_q=0_symmetric}
H_{\text{s},q\sim 0}=\int \eta_{s}(x)\partial_x\phi_\rho(x) dx
\end{equation}
It is clear that this part of the disorder can be eliminated by
the transformation $\phi_\rho(x) \to \phi_\rho(x) +\int^x\frac{\pi
K_{\rho+}}{u_{\rho+}} \eta_s(x')dx'$
For the $q \sim 0$ part of the antisymmetric random potential, we obtain~:
\begin{equation}
H_{\text{a},q\sim 0}=\int dx \eta_a(x) \sum_\sigma \left[
\psi_{R,0,\sigma}^\dagger \psi_{R,\pi,\sigma}+
\psi_{L,0,\sigma}^\dagger \psi_{L,\pi,\sigma}+ \text{ H. c.} \right]
\end{equation}
The bosonized form of that operator is the following:
\begin{equation}
H_{\text{a},q\sim 0}=\int dx \frac{\eta_a(x)}{\pi \alpha}
\left[
e^{\imath(\phi_{\rho-}+\theta_{\rho-})} \cos(\phi_{\sigma -}+\theta_{\sigma-})
+e^{\imath(-\phi_{\rho-}+\theta_{\rho-})} \cos(\phi_{\sigma-}-
\theta_{\sigma-})
+ \text{ H. c.} \right]
\end{equation}
 From that equation, we see that the $q \sim 0$ part of the
antisymmetric disorder is not coupled to the gapless charge symmetric mode.
Moreover, it always contain one term that has exponentially decaying
correlations. Therefore, it cannot break any gap by an effect {\em \`a la} Imry
Ma and cannot generate any relevant term by a massive mode integration.
It will thus be possible to drop it safely in the following.
Then, we have to consider the $2k_F$ part of the disorder.
We have for the $2k_F$ coupling to disorder two terms~:
\begin {eqnarray} \label {picdw}
H_{a} & = & \int\xi_{a}(x)O_{CDW^{\pi}}(x) + \xi_{a}^{*}(x)
O_{CDW^{\pi}}^{\dagger}(x) dx \\
\label{zerocdw}
H_{s} & = & \int\xi_{s}(x)O_{CDW^{o}}(x) +
\xi_{s}^{*}(x)O_{CDW^{o}}^{\dagger}(x) dx
\end{eqnarray}
Where $\overline{\xi_{n}(x)\xi_{n'}(x')^{*}}=D_{n}\delta_{n,n'}\delta(x-x')
(n,n'=a,s)$, the $\xi_{n}$ being random Gaussian distributed potentials.
The operators $O_{CDW^o}$ represents  the in-phase
charge density wave, and $O_{CDW^{\pi}}$ the out of phase one.

As before we assume that the disorder
weak enough not to destroy the gaps in the system.
We have already argued that the $q \sim 0$ is irrelevant to our problem.
Concerning the $2k_F$ part, we only retain the massless mode.
The situation is quite similar to the one
of a XXZ spin chain in a random magnetic field. A XXZ spin chain is a
Hubbard chain at half filling, and thus has a charge gap. The random
magnetic field couples to the spin density that contains (frozen)
charge degrees of freedom. However, the random magnetic field only
affects the spin degrees of freedom and does not break the charge
gap. By analogy, we  expect that even when the random potential gets
relevant it will not break the spin gap or
the gap in the antisymmetric charge  mode.
Since the gaps are stable, we can obtain simplified forms for the
couplings by replacing the fields by their mean values as we did in
the spinless fermions problem.

\subsubsection{$SC^d$  sector} \label{localise_scd}

We want to analyze the effect of the weak random potential introduced
through (\ref{picdw}-\ref{zerocdw}). Making use of the full
expressions of $O_{CDW^{o,\pi}}$ and replacing the gapped fields by
their mean values (see sector I of table \ref{table1}) we obtain the
following simplified forms~:
\begin{eqnarray}
\label{simpl_repuls}
O_{CDW^{o}} & \sim & e^{\imath \phi_{\rho +}} \sin(\phi_{\rho -}) \\
O_{CDW^{\pi}} & \sim & e^{\imath \phi_{\rho +}} \sin(\theta_{\sigma-})\cos
\phi_{\sigma+}
\end{eqnarray}
These two operators have exponentially decaying
correlation functions and no direct coupling with disorder would exist
if one just took into account the mean values of the fields
$\phi_{\rho,-}$ and $\theta_{\sigma,-}$. As in the spinless case one
should integrate over fluctuations to get the effective coupling
\begin{equation}
\label{ougl}
S_{\rho +}^{\text{disorder}}=\int \xi_{\text{eff.}}(x)e^{\imath
2\phi_{\rho +}(x,\tau)} dx d\tau + \text{H. c.}
\end{equation}
(\ref{ougl}) can be viewed as
the coupling of the fermions with the $2(k_{Fo} \pm k_{F\pi})$ Fourier
component of the disordered potential i.~e. to a $4k_F$ charge density
wave. The origin for such a $4k_F$ charge density wave can be
understood in simple terms: at half filling, the strong on site
repulsion puts one fermion per site, meaning that there are no $2k_F$
CDW fluctuations.
However, the fermion density is maximum on the lattice site and
minimums in between giving the $4k_F$ charge density wave
fluctuations. In addition due to the spin gap occuring in a ladder with
an even number of legs there are no $2k_F$ fluctuations in
the spin density as well.
As we move away from half filling, the spin gap will survive
as well as the absence of $2k_F$ fluctuations. Therefore, a random
potential can only couple to the $4k_F$ component of the fermion density
even away from half
filling. This is to be contrasted to the case of a single chain where
the dominant coupling occurs through the $2k_F$ charge fluctuation.
One thus expects the disorder effects to be weaker in the ladder system.
One can also recover directly the
$4k_F$ CDW  by looking at higher Fourier components of the density
in the bosonization formulas. The physics of the metal insulator
transition can be here interpreted as the pinning-depinning transition
of this $4k_F$ charge density wave.

Due to the presence of the gaps,
the problem has in fact been formally reduced to a problem of one chain
of spinless fermions with disorder .
Using the results from the one chain problem we find
that the localization-delocalization occurs at $K_{\rho +}=3/2$.
Since purely
repulsive interaction imply $K_{\rho+}<1$ the $d$-wave phase is therefore
unstable to arbitrarily weak disorder.
The symmetric (\ref{zerocdw}) and the
antisymmetric (\ref{picdw}) part of the disorder contribute
equally to destroy the d-wave superconductivity, in contrast with the
spinless case where the antisymmetric part was the most relevant.
The localization length of the 2 chain system with spin and purely repulsive
interactions can be obtained by a similar method than for the spinless
case  and is
\begin{equation}  \label{locle2ch}
\frac{L_{\text{2 ch.}}}{\alpha} \sim \left(\frac{v_F^2}{D\alpha}\right)^{2/(3-2K_{\rho +})}
\end{equation}
and therefore longer than the
corresponding one for one chain with
repulsive interactions \cite{suzumura_scha,giamarchi_loc}
\begin{equation} \label{singrep}
\frac{L_{\text{1 ch.}}}{\alpha}\sim \left(\frac{v_F^2}{D \alpha}\right)^{1/(2-K_{\rho +})}
\end{equation}
As for the spinless case
(\ref{locle2ch}) is applicable if one is far enough from the
noninteracting point so that disorder does not destroy the gaps created
by the interactions. In that case one sees from (\ref{locle2ch}) that
there is a considerable delocalization in the ladder. Indeed for weakly
repulsive interactions $K_{\rho +}\sim 1$, the localization length
becomes much longer than the mean free path $l$, since $L_{\text{2 ch.}}\sim
\alpha (l/\alpha)^2$, instead of $L \sim l$ for a single chain.
However the more repulsive the interactions become, the more the system
localizes (one recovers $L_{\text{2 ch.}}\sim l$ for $K=1/2$).

The temperature dependence of the conductivity can be obtained above the
pinning temperature $T_{\text{pin.}}=\frac{u_{\rho+}}{L_{\text{2ch.}}}$
(see appendix~\ref{condumem}). One gets

\begin{equation} \label{dcond}
\sigma(T)\propto T^{2-2K_{\rho+}}
\end{equation}
For $K_{\rho+}<1$, the conductivity {\bf decreases} as $T \to 0$ even for
temperatures much higher than  $T_{\text{pin.}}$.
There is no remnant of the ``superconducting'' behavior of the pure
system in the whole $SC^d$ sector ($K_{\rho+}<1$).

\subsubsection{$SC^s$ sector} \label{lasupras}

For sector III, the $O$ operators take a different simplified
form, due to the different gaps in the system
\begin{eqnarray}
\label{simpl-o-attrac}
O_{CDW^{o}} & \sim &  e^{\imath\phi_{\rho +}}\cos(\phi_{\rho -}) \\
\label {simpl-pi-attrac}
O_{CDW^{\pi}} & \sim & e^{\imath\phi_{\rho +}} \sin(\theta_{\sigma -})
\end{eqnarray}
By substituting in
(\ref{picdw}) and (\ref{zerocdw}) and integrating over fluctuations
we end with an action of
the form (\ref{ougl}). This time, $K_{\rho +}>1$, so
the localization-delocalization transition can be reached at
$K_{\rho +}=3/2$. This transition arises for much weaker attraction
than in the one dimensional case \cite{giamarchi_loc}
where $K_{\rho}=3$. This critical value of $K$ can be realized for a
simple Hubbard model
(the maximum $K$ for the Hubbard model is $K=2$
\cite{kawakami_bethe_U<0,giamarchi_persistent_1d})
whereas the one chain Hubbard model is always localized even for very
negative $U$ \cite{giamarchi_persistent_1d}.
In addition the localization length is increased
\begin{equation}  \label{locat}
 \frac{L_{\text{2 ch.}}}{\alpha}= \left(\frac{v_F^2}{D\alpha}
\right)^{\frac{2}{3-2K_{\rho +}}}
\end{equation}
whereas in the one chain case
\begin{equation} \label{singat}
\frac{L_{\text{1 ch.}}}{\alpha}=(\frac{v_F^2}{D\alpha})^{\frac{1}{3-K_{\rho}}}
\end{equation}
Note that here the localization length has the same dependence in
disorder on the attractive (\ref{locat}) and the repulsive (\ref{locle2ch})
side, whereas for a single chain the localization length is {\bf
reduced} on the attractive side due to the formation of a spin gap
(compare (\ref{singat}) and (\ref{singrep})). For the ladder this come
from the fact that both in the attractive and repulsive sector, three of
the modes are always gapped.

The conductivity above the pinning temperature behaves as
\begin{equation} \label{laconds}
\sigma(T)\sim T^{2-2K_{\rho+}}
\end{equation}
with again the same exponent than in the d-wave sector (\ref{dcond}).
However, since now $K_{\rho +}>1$ the conductivity now {\bf decreases}
with decreasing $T$. There
will thus be for $1<K_{\rho +}<3/2$ a maximum in the conductivity for
$T\sim T_{\text{pin.}}$, and the resistivity will go to zero for high
values of $K_{\rho +}$. This maximum can be seen as a remnant of the
superconducting behavior of the pure system.
For $K_{\rho+}>3/2$, the system has infinite conductivity for $T\to
0$.

\subsubsection{$CDW^\pi$ sector }

Let us now consider sector IV. In that sector, one has strong fluctuations
towards a $CDW^\pi$ phase. Such phase is the analog of the $CDW^\pi$
that existed in the spinless fermion problem.
We see that the coupling to disorder reduces to (see table \ref{table1})
\begin{equation}
\int dx \xi_a(x) e^{\imath \phi_{\rho+}} +\text{H. c.}
\end{equation}
As in the spinless fermion case that antisymmetric Charge density Wave
only couples
to the antisymmetric disorder.
The RG equation for disorder is
\begin{equation}\label{cdwpi_2ch_spin_RG}
\frac{dD_a}{dl}=\left(3-\frac{K_{\rho+}}{2}\right)D_a(l)
\end{equation}
The antisymmetric disorder is thus relevant for $K_{\rho+}<6$.
Since the $CDW^\pi$ phase only exists at $K_{\rho+}<1$ the $CDW^\pi$
is always very strongly pinned by disorder.
Using (\ref{cdwpi_2ch_spin_RG}) we obtain for the localization length in
that phase
\begin{equation}
\frac{L_{\text{loc.} ,CDW^\pi}}{\alpha} \sim \left(\frac {v_F^2} {D_a\alpha}
\right)^{\frac{2}{6-K_\rho+}}
\end{equation}
In the classical limit $K_{\rho+} \to 0$ one recovers again the standard
result \cite{fukuyama_pinning}
for the pinning of a classical CDW.

The conductivity of the $CDW^\pi$ above the pinning temperature
behaves as
\begin{equation}
 \sigma(T)\sim T^{2-\frac{K_{\rho+}}{2}}
\end{equation}
showing since  $K_{\rho+}<1$  a very rapid decrease
in the conductivity as $T \to 0$. This behavior is a consequence of the
very strong pinning of the $CDW^\pi$.
This feature of the antisymmetric CDW is similar to the one occuring for
the spinless ladder.

\subsubsection{OAF sector}

In the case of the orbital antiferromagnet,
the coupling to disorder is made of 2 terms:
One term comes from $O_{CDW^o}$ the other one from $O_{CDW^\pi}$.
According to the preceding sections, these terms contain respectively
$\cos \phi_{\rho-}$ and $\cos \phi_{\sigma+}\sin\theta_{\sigma-}$
and (see table \ref{table1}) therefore have exponentially decaying fluctuations.
In order to get nontrivial results,  the
massive modes have to be integrated out as in the preceding sections .
This again leads to an action of the form (\ref{ougl}) and
the disorder in the OAF phase is relevant for $K_{\rho+}<3/2$.
The OAF phase is therefore {\bf as delocalized} as the superconducting
$SC^s$ phase, although the pure system does not exhibit nay obvious
superconducting order parameter.
The localization length in the disordered OAF is
\begin{equation}  \label{locoaf}
\frac{L_{loc.}}{\alpha}=\left(\frac{v_F^2}{D \alpha}\right)^{\frac{2}{3-2K_{\rho+}}}
\end{equation}
For $K_{\rho+}>3/2$ we have a metallic phase.

The disorder leads to a conductivity  of the form
\begin{equation} \label{condoaf}
\sigma(T)\propto T^{2-2K_{\rho+}}
\end{equation}
The conductivity in the OAF is therefore identical, as far as the
temperature dependence is concerned, as the one in the $SC^s$. It will
exhibit in the localized phase $1<K_{\rho+}<3/2$ the same
maximum in the conductivity for $T\sim T_{\text{pin.}}$.
Once again one sees that the transport
properties can hardly be guessed from the phase diagram of the pure
system. The OAF is thus also an excellent candidate for a
``superconducting'' behavior.

Using (\ref{g1g2=fUV}), it is possible to get some hints on the
parameter regime of the extended Hubbard model in which the OAF could
be achieved. One is in the OAF sector if $g_1>0$ and
$K_{\rho+}>1$. In the extended Hubbard language it means
\begin{equation}
2V(2-\cos(2k_Fa))<-U<2V\cos(2k_Fa)
\end{equation}
Let us assumes a local repulsion $U>0$ and that one is
close to half filling $\cos(2k_Fa)\sim -1$. In that case one reaches the
OAF for moderate nearest neighbor attraction $V < -U/6$. Such a
situation is likely enough to be realized, specially if additional
attractive mechanism such as phonons are taken into account.

\subsubsection{Differences with the spinless ladder}

\begin{figure}
\centerline{\epsfig{file=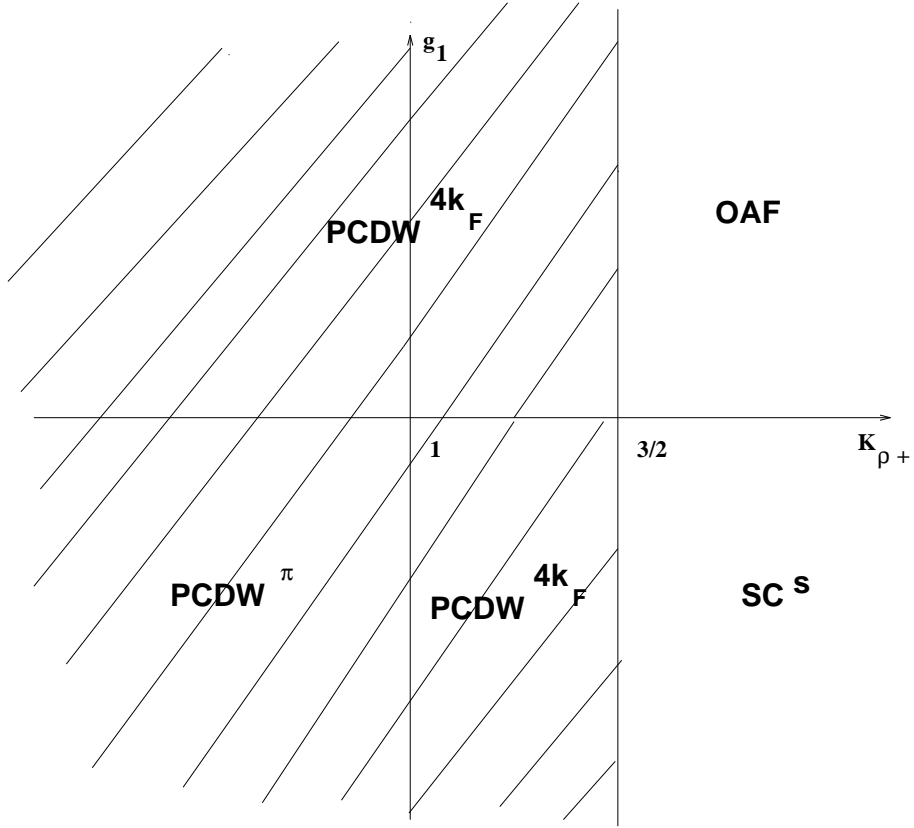,angle=-90,width=12cm}}
   \vspace{0.5cm}
\caption{The phase diagram of the disordered 2 chain Hubbard model in
terms of $g_1$ and $K_{\rho+}$. The $Sc^d$ phase is completely eaten
by the $PCDW^{4k_F}$ phase whereas the OAF and the $SC^s$ persist if
there is enough attraction. Delocalization occurs for $K>3/2$ i.e. for
less attractive interactions than in the one chain case.}
\label{dis_hubb}
\end{figure}

The spinless ladder and the ladder with spin show some marked physical
differences . Some of them are due to the fact that interchain hopping
has a different impact on fermions with spin and spinless fermions.
In a system of spinless fermions, energy can be gained from hopping
only if one site of the rung is occupied and the other one is empty
due to the Pauli principle. This induces an enhancement of density
fluctuations. On the other hand, a system with spin can gain energy
from interchain hopping by having the two sites of the rung occupied
by fermions of opposite spins. This leads to spin gap formation and a
\emph{smoothening} of density fluctuations.
This effect is enhanced in the presence of a purely
repulsive interaction as it tends to smoothen the density fluctuations in a
system with spins, whereas it enhances them in a spinless system
\cite{giamarchi_persistent_1d}.
This has already important consequences in the pure case. In
particular, the positions of the $SC^d$ and OAF phases are different
(see figure~\ref{pure_tV} and
figure~\ref{pure_hubbard}) as
the d-wave in the spinless system needs some
amount of attraction whereas it is achieved from completely
repulsive interactions in the ladder with spin. In the presence of
disorder, the d-wave phase of the spinless system can be stabilized by
sufficiently attractive interactions, whereas in the system with spin
it is {\bf always} unstable(see figures~\ref{disordered_tV} and \ref{dis_hubb} ), being replaced by an s-wave
superconducting phase for attractive interactions.
 Also, in the presence of disorder, the  system with spin due to the
smoothening of the density, shows delocalization compared to the one chain case {\bf both}
for the attractive and the repulsive side. On the other hand for the
spinless system, the
reinforcement of  the density fluctuations,  enhances localization on
the repulsive side. The attractive side on the other hand is totally
delocalized.

In both case the s-wave phase (occuring for
attractive interactions) is very strongly stabilized by the interchain
hopping. This can be understood by a picture of tightly bound pairs
that behave in both cases as hard core bosons. In that case, the
statistics do not influence any more qualitatively the transport properties.
 Similarly both systems tend to form charge
density waves that are extremely well pinned by disorder (usually much more
easily than their one chain counterpart). In the case of fermions with
spin, this requires some mixing of attractive and repulsive
interactions so that pair of fermions of opposite spins are formed in
the chains. These pairs then have hard core bosons interactions so
that the situation becomes analog to the spinless fermions case. This
explains the enhancement of pinning for the antisymmetric charge
density wave phase. However, in the system with spin with purely
repulsive interactions there is \emph{no} $CDW^\pi$ in contrast with
the spinless system.
Both systems also present an OAF phase
that reveals quite stable in the presence of a small disorder.
For the spinless ladder the OAF is even stable close to the
noninteracting point.
Finally, an interesting similarity between the system of fermions with
spin and the system of spinless fermions is that  pinning on {\bf
two} different CDW phases are possible depending on the interactions:
either the antisymmetric $2k_F$ CDW or a
$4k_F$ CDW. In these two localized phases the behavior of the conductivity
at high frequency or high temperature and of the localization length at
small disorder are very different (the difference appears in the
exponents) the $4k_F$ being much less well pinned than the $2k_F$.
This is to be contrasted to the one chain case where only one
pinned charge density wave phase is realized. Therefore, we may expect to see,
for weak disorder,
a crossover between two different pinned charge density wave phases in
the two chain system when varying the strength of the interactions.
Such a crossover needs a more detailed study. Unfortunately it cannot be
tackled by the RG since it occurs deep in the localized regime. One
interesting question is whether such a transition still occurs for strong
disorder.

\subsection{Transport properties}\label{transport}

The ladder with spin shows therefore in presence of disorder transport
properties drastically different from the one one could naively expect
form the pure phase diagram. In particular the d-wave phase disappears
and does not exhibit any remarkable conductivity. Let us look in more
details in the transport properties and compare them to what happens
in a single chain\cite{giamarchi_loc} for the various sectors.

\subsubsection{Conductivity}\label{conductivity_2ch_w_spin}

\begin{figure}
 \centerline{\epsfig{file=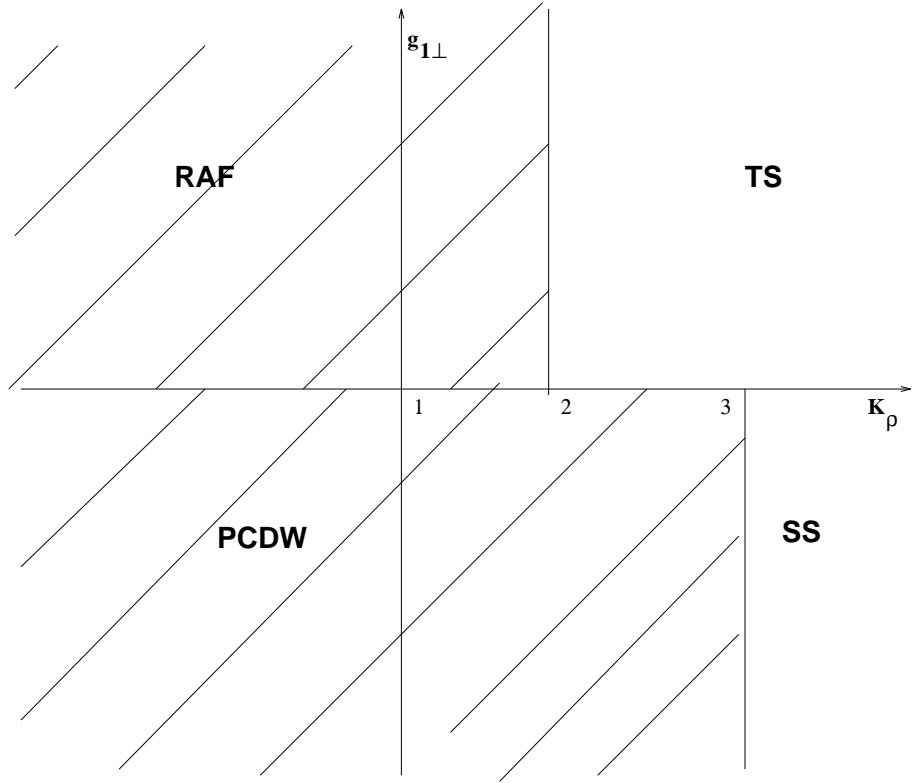,angle=-90,width=12cm}}
   \vspace{0.5cm}
\caption{The phase diagram of the disordered 1 chain Hubbard model in
terms of $g_{1\perp}$ and $K_{\rho}$. Delocalization occurs for
$K_\rho>3$ for $g_{1\perp}<0$ and for $K_\rho>2$ for $g_{1\perp}>0$.}
\label{1ch_hub_dis}
\end{figure}

As was mentioned in section~\ref{lasupras}, the ladder s-wave phase is
much more stable to disorder than its one chain counterpart (see
figures~\ref{1ch_hub_dis} and~\ref{dis_hubb}). This effect
manifest itself in the location of the superconducting-localized
transition, and in the localization length.
As for the spinless case, this effect is entirely controlled by the
interactions and going from one to two chains affects the {\em power
law dependence} of the localization length with disorder. It is thus
much stronger than the increase of localization length occuring for a
noninteracting system (proportional to the number of channels).
In presence of interactions the behavior of the localization length
cannot be guessed by analogies with the non-interacting system.
The case of spinless fermions where repulsive interactions make the two
chain system more localized than the one chain system is an excellent
counter-example.

\begin{figure}
\centerline{\epsfig{file=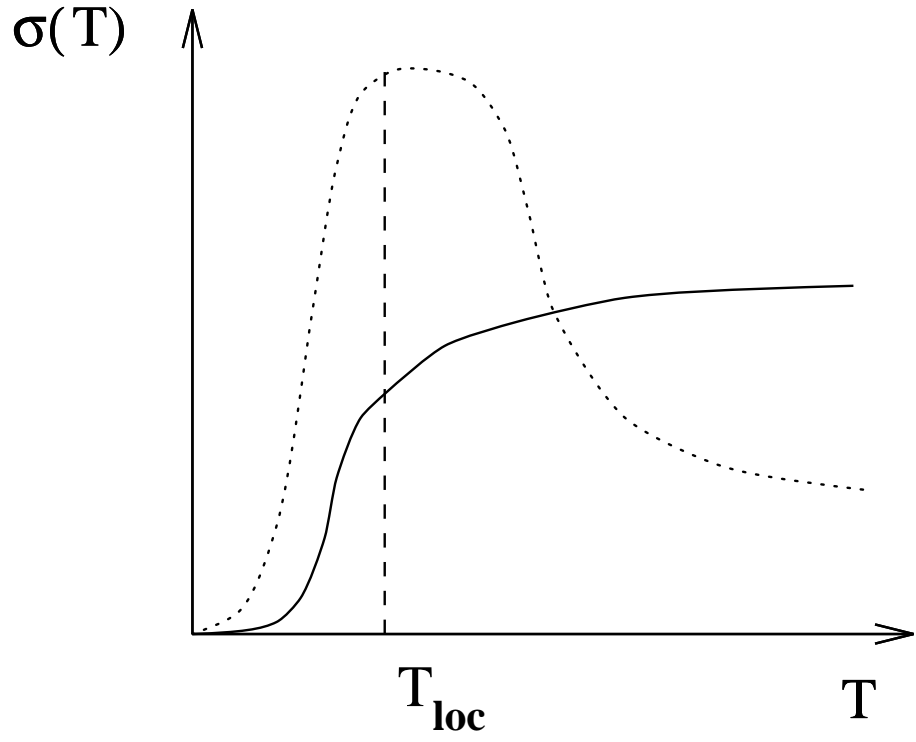,angle=-90,width=12cm}}
\vspace{0.5cm}
\caption{Behavior of the conductivities of the s-wave (dotted line)
and d-wave (solid line) superconductor as a function of temperature.
For $T\gg T_{\text{loc.}}$, $\sigma(T)\propto T^{2-2K_\rho+}$.
For the d-wave, there is no maximum in the conductivity and therefore
no remnant of superconductivity in the localized phase.}
\label{sigma_vs_T}
\end{figure}

The resistivity (see (\ref{laconds})) is also dropping much faster than
for one chain for which $\sigma_{\text{1ch.}}(T)\sim T^{2-K_\rho}$.
The ladder is thus a much better conductor than a single chain both
because of the scale of localization and because of the better
temperature dependence. In addition even in the localized phase the
conductivity will increase for all values of $K_{\rho+}$ for which
the s-wave phase exists in the pure system, till one reaches the
localization temperature $T_{\text{pin}}$.This behavior is qualitatively
sketched on figure \ref{sigma_vs_T}. The $SC^s$ phase shows
therefore all the ``good'' characteristics of a ``superconducting''
phase, and in that respect is much more normal than its one-chain
counterpart.

For repulsive interactions a different physical situation
occurs. The system is still less localized than the one chain
counterpart. The transition occurs for a smaller value of $K_{\rho+}=3/2$
(versus $K_\rho=2$ for a single chain), and the localization length is
larger than for one chain (see (\ref{locle2ch})). Contrarily to the
single chain where the pinned phase is a random antiferromagnet,
here the presence of the spin gap forces the localized phase to become a
pinned $4k_F$ CDW. However the $SC^d$ phase is completely wiped out by
the disorder, and what is more surprising, no trace of this
``superconducting'' phase can be found in the high temperature ($T >
T_{\text{pin}}$) of the conductivity (see (\ref{laconds})). In
particular $\sigma(T)$ decreases monotonically even at high
temperature in stark contrast with the $SC^s$ phase as shown on
figure~\ref{sigma_vs_T}.
This again illustrate the fact that the transport properties are not
linked to the behavior of the superconducting order parameter but to the
{\bf density} fluctuations. For a single chain since the density
exponent and the superconducting one are related by
$K_{\text{density}}\sim 1/K_{\text{supra}}$ when superconducting
fluctuation increases density fluctuations necessarily decrease and the
system becomes a better conductor. Or course this is also true in
presence of a true superconducting order in higher dimensional systems.
For the single chain the fact that superconducting fluctuations do not
necessarily imply better transport also appears one the fact that the
attractive Hubbard model is {\bf more} localized than the repulsive one
\cite{giamarchi_persistent_1d}: when the interactions go from repulsive
to attractive a spin gap opens and the density fluctuations are
suddenly lowered making the system more easy to pin.
Similar effect occurs in the d-wave phase of the ladder, in a more
dramatic way: the d-wave phase do not look superconducting at all since
it leaves enough room for enough $4k_F$ charge fluctuations.
Note that the more repulsive the interactions will be the worse the
conductivity, in a similar way than for the single chain where the phase
is a spin density wave. The interchain hopping has thus two effects~:
on the one hand it leads to the appearance of the spin gap that wipes
the SDW and replaces it by the $SC^d$ wave and
on the other hand it freezes the density
fluctuations (in particular the transverse charge modes).
Those gaps suppress $2k_F$ CDW fluctuations, and localization happens
only through coupling to $4k_F$ CDW fluctuations. Since the mechanism
for  localization is the same for all sign of the interactions, the
transport properties are only weakly dependent of the sign of the
interactions.  This charge freezing is the
dominant effect on transport. The two effects are essentially unrelated.

The most remarkable phase is the OAF  which is an
illustration of the above. This phase has a localization length and a
$\sigma(T)$  {\bf as good as} a genuine $SC^s$ wave phase ! and yet has
no genuine superconducting  order parameter. In fact the absence of order
parameter is
here also due to the spin gap since for a single chain the corresponding
phase is a triplet superconducting phase. However the fact that density
fluctuations are already very small in this phase remains (and is helped
by the freezing of transverse charge fluctuations), giving the
remarkable transport properties of this phase. This remarkable property
is not an artifact of the potential scattering and persists even if
coupling to different form of disorder is included. In
particular the
superconducting-like transport properties of the OAF also exist in the
presence of a random hopping along the chains and a random interchain
hopping amplitude (see appendix \ref{other_perturbations}).
Note that this phase has analogies of the so-called flux phase
\cite{affleck_marston,kotliar_fluxphases,anderson_fluxphases,lederer_fluxphases},
the size of the plaquette is
here fixed by the interparticle distance, and of course this phase could
not be reached
for a pure Hubbard model (at the opposite of what was claimed in higher
dimension). Whether for such a phase, a sort of Meissner effect also
exists is of course a very interesting question. The connection
between the one dimensional antiferromagnet and its 2d or 3d
counterparts clearly deserves further investigations. In particular
in two dimensions a phase offering some similarities
with the one dimensional OAF, has been proposed for the high-Tc
superconductors \cite{varma_review}.

\subsubsection{Persistent currents}

In a similar way than for the spinless case one can compute the charge
stiffness. For the ladder with attractive interactions one has for
$K_{\rho+},K_\rho<3/2$
\begin{equation}
{\mathcal D}(L)={\mathcal D}(0)-(\frac{L}{L_{\text{loc.,1 ch.}}})^{3-K_\rho}
\end{equation}
whereas for a two chain one, it is
\begin{equation}
{\mathcal D}(L)={\mathcal D}(0)-(\frac{L}{L_{\text{loc.,1
ch.}}})^{3-2K_\rho+}
\end{equation}
These formulas are valid for $\alpha \ll L \ll L_{\text{loc.}}$.
It is easy to see that they lead to a smaller reduction of the conductivity
stiffness in the 2 chain case, in agreement with the fact that
$L_{\text{2 ch.}}>L_{1ch}$.
For repulsive interactions, it is of the form
\begin{equation}
{\mathcal D}(L)={\mathcal D}(0)-(\frac{L}{L_{\text{loc.,1 ch.}}})^{3-K_\rho}
\end{equation}
Whereas for the 2 chain case, it is of the form~:
\begin{equation}
{\mathcal D}(L)={\mathcal D}(0)- (\frac{L}{L_{\text{loc.,1ch.}}})^{3-2K_\rho+}
\end{equation}
and the 2 chain system has a smaller reduction of conductivity
stiffness than the one chain system.
So up to prefactors the reduction in stiffness in the ladder system with
spins is identical for repulsive and attractive interactions and the
reduction of conductivity stiffness also
shows no abrupt change as one goes from attractive to repulsive
interactions.
By contrast, in the one chain case, attractive interactions induce a spin gap and
localization arises from coupling of a single massless mode to $2k_F$ disorder.
This gap closes for repulsive interactions
and localization arises from the coupling of 2 massless modes with
the $2k_F$ random potential. This causes the abrupt change in
transport properties and
charge stiffness\cite{giamarchi_loc,giamarchi_persistent_1d}
when one goes from attractive to repulsive interactions.
This is related to the fact that the localization lengths for attractive
and repulsive interactions have the same dependence on disorder, in
marked contrast both with the spinless problem and the single chain with
spins. The effect of increase of persistent current by repulsive
interactions occuring in the single chain \cite{giamarchi_persistent_1d}
is thus either absent or strongly reduced (not an exponent effect any
more) in the ladder. It would of course be interesting to investigate in
ladder with more than two legs to see if this effect reappears and check
for possible difference of behavior between odd and even legs ladders.

\section{Coupled ladders}
\label{MEANFIELD}

\subsection{Mean field treatment}  \label{mfsimple}

In the preceding sections, we have been considering isolated bichains.
To describe realistic compounds, such as SrCuO, and have a finite
temperature phase transition, interchain coupling should be taken into
account. A realistic coupling is of course single particle hopping
between the ladders. However in ladders, due to the existence of single
particle gaps (spin and antisymmetric charge mode) for the ladder,
single particle hopping is irrelevant, provided that the inter-ladder
hopping is much smaller than the gaps of the system. One has therefore
to consider only the  particle-hole (or particle-particle)
coupling  generated by the single particle hopping
\cite{brazovskii_transhop}.
Such couplings can lead to an ordered phase at a finite temperature.
As is very reasonable on physical grounds such interchain couplings
stabilize the dominant one-dimensional instability.
We focus here on the existence of a stable d-wave superconducting
phase. This allows us to keep only the particle-particle (or Josephson)
coupling between the ladders.
The  Hamiltonian for the coupled ladders system is:
\begin{equation}
\label{array_of_bichains}
H=\sum_{n}\left[H_{\text{disordered 2 chain system},n}
+\frac{J}{2}\int dx (O_{SC,n}^\dagger(x)O_{SC,n+1}(x)+
O_{SC,n+1}^\dagger(x)O_{SC,n}(x)) \right]
\end{equation}
Where $O_{SC,n}$ is the operator for (d-wave or s-wave) superconductivity
for the $n-$th ladder and $J$ is the strength of the Josephson coupling.
On can simplify further the Hamiltonian (\ref{array_of_bichains}) by
keeping only massless modes in the ladder. Doing so we assume
that the spin gap and the interchain gap of the two
chain system are much larger than the disorder and much larger than
the Josephson coupling. However, we make no assumption on the relative
magnitude of the Josephson coupling and the strength of the random potential.
The resulting Hamiltonian is, both for the case where the dominant
instability is s-wave or d-wave superconductivity
\begin{eqnarray}
\label{simplified_array}
H & = & \sum_{n} (\int \frac{dx}{2\pi} \left[
u_{\rho+}K_{\rho+}(\pi\Pi_{\rho+,n})^2 +
\frac{u_{\rho+}}{K_{\rho+}}(\partial_x\phi_{\rho+,n})^2 \right] +
\int \frac{dx}{\pi\alpha} \left[ \xi_{\text{eff.},n}(x) e^{\imath
2\phi_{\rho+,n}
}+ \text{H. c.} \right] \nonumber \\
& &+J\int dx \cos(\theta_{\rho+,n}-\theta_{\rho+,n+1}))
\end{eqnarray}
To solve (\ref{simplified_array}) we treat the Josephson coupling in
mean field assuming the existence a finite superconducting order
parameters $\langle \cos(\theta_{\rho+})\rangle$.
By making the replacement
 $\cos(\theta_{\rho+,n}-\theta_{\rho+,n+1}) \to \langle
\cos(\theta_{\rho+})\rangle \cos(\theta_{\rho+,n})$, the Hamiltonian
(\ref{simplified_array}) becomes
the one of an  isolated ladder system in an external field,
the value of which is determined by a self-consistency condition.
The Hamiltonian is then
\begin{eqnarray}
\label{mean-field-SC}
H_{MF} & = & \int \frac{dx}{2\pi} \left[ u_{\rho+}K_{\rho+}(\pi\Pi_{\rho+})^2 +
\frac{u_{\rho+}}{K_{\rho+}}(\partial_x\phi_{\rho+})^2 \right] +
\int \frac{dx}{\pi\alpha} \left[ \xi_{\text{eff.}}(x) e^{\imath 2\phi_{\rho+}
}+ \text{H. c.} \right] \nonumber \\
& & -\frac{W}{(2\pi\alpha)^2}\int dx \cos(\theta_{\rho+})
\end{eqnarray}
with the self-consistency condition $W=J\langle \cos(\theta_{\rho+}) \rangle$.

\begin{figure}
\centerline{\epsfig{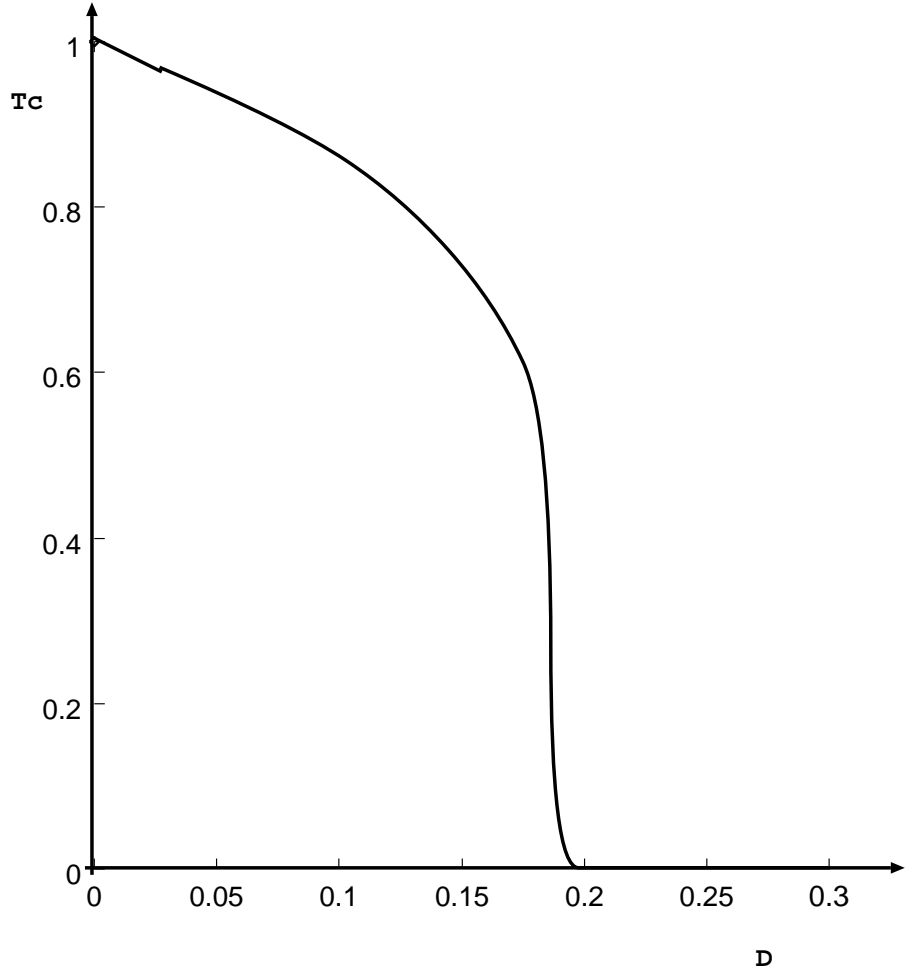}}
   \vspace{0.5cm}
\caption{Tc as a function of disorder for the d-wave phase ($K_\rho
=0.5$). Tc drops quickly to zero for $D\simeq 0.2$ }
\label{TcD_dwave}
\end{figure}

\begin{figure}
 \centerline{\epsfig{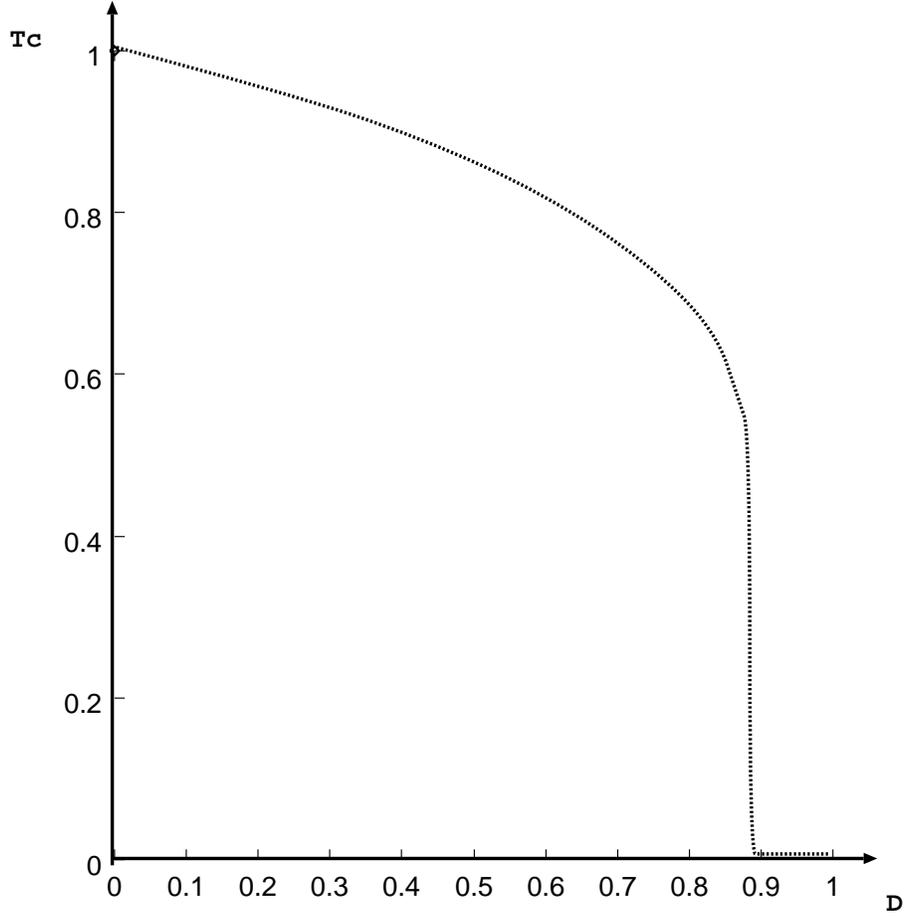}}
   \vspace{0.5cm}
\caption{Tc as a function of disorder for the s-wave phase ($K_\rho
=1.2$) . Tc drops to zero for $D\simeq 0.9$. Note the initial linear
decay of Tc that shows that Anderson Theorem does not hold in coupled
chain system due to strong localization effects and absence of a
diffusive regime.}
\label{TcD_swave}
\end{figure}

The equation determining $T_c$ is
\begin{equation}\label{MF_general}
\frac{1}{J}=\frac{1}{(2\pi\alpha)^2}\int dx \int_{0}^{\beta_c} d\tau \langle
T_{\tau}
\cos\theta_{\rho+}(x,\tau) \cos\theta(0,0) \rangle_{H_0}
\end{equation}
  with $\beta_c=\frac{1}{T_c}$ and $H_0$ is $H_{MF}$ for $W=0$.
To solve (\ref{MF_general}), one has to compute the finite
temperature superconducting response function of a ladder in the
presence of disorder. There are presently no methods  to do this
exactly, but one can get an accurate solution for $T_c$ by making
some simplifying approximations \cite{suzumura_mean_field}.
First, one notices that
a finite temperature induces a cutoff length $l(T)=u_{\rho+}/T$ beyond
which all correlation functions decay exponentially to zero.
We make thus the
approximation that beyond $l(T)$ all correlation functions are truly
zero and below $l(T)$ they are equal to the $T=0$ correlation functions.
This allows us to use the RG equations introduced in section~\ref{2ch_Hubbard}.
If we denote by $\chi$
the superconducting response function, when we change the running
cutoff $\alpha(l) \to \alpha(l) e^{dl}$ we have $\chi \to \chi
\exp(-\frac{dl}{2K(l)})$. Thus to compute correlation functions at
lengthscale $R$ it is sufficient to integrate the RG equation from the
cutoff up to $R$ and follow the renormalization of the response
function.
Making use of these two approximations, the equation
giving  $T_c$ simplifies into~:
\begin{equation}
\label{MF_simple}
\frac{1}{J}=\int_{\alpha}^{u_{\rho+}/T}\frac{R dR}{2\pi\alpha^2}
\exp(-\int_{0}^{\ln(R/\alpha)}\frac{dl}{2K(l)})
\end{equation}
The values of $K(l)$ are obtained by numerically solving the RG equations:
\begin{eqnarray}
\frac{dK}{dl}=-D(l) K(l)^2 \\
\frac{dD}{dl}=(3-2K(l))D(l)
\end{eqnarray}
the values of $T_c$ for$ K=0.5, 1.2$ and $J=0.1$ as
a function of $D$ are shown on figures \ref{TcD_dwave}  and
\ref{TcD_swave} respectively.
We note that for $K=1.2$ we have an s-wave superconducting phase and
for $K=0.5$, a d-wave phase. This can be expected since the interchain
coupling stabilizes the dominant one dimensional fluctuation
(see figure~\ref{dis_hubb}).
We see that (see fig. \ref{TcD_swave}) as in the case of the
single chain mean field theory\cite{suzumura_mean_field} of
superconductivity we have an initial linear decrease of the critical
temperature with disorder strength.
This is to be contrasted with the standard
mean field theory of the s-wave superconductor in three dimensions
being based on a diffusion approximation that
does not include Anderson localization effects
and gives $T_c$ independent
of the disorder. This is the well known Anderson theorem. The linear
decrease of $T_c$ with
the strength of disorder for s-wave superconductivity in our chain
mean field is due to localization effects. This peculiar situation is
due to the
absence of a diffusive regime in one dimensional disordered systems,
which implies that their response functions are always affected by
localization effects.

For a d-wave superconductor one expects in mean-field theory a
linear decrease of $T_c$ as a function of $D$ (see e.g.
Ref.~\onlinecite{millis_mean_field}). For the ladder system however the
decrease of $T_c$ is mainly due to the localization effects, similarly
to the s-wave superconductor case. Although it
indeed starts linearly for small disorder (see figure~\ref{TcD_dwave}),
localization effects manifest themselves by the the sudden drop to
$T_c=0$ at a critical disorder strength (see
fig. \ref{TcD_swave},\ref{TcD_dwave}).

For identical Josephson coupling between the bichains, the
critical disorder strength is smaller for the d-wave superconductor
than for the s-wave one.

\subsection{Simplified Treatment:}

Although the mean field theory allows an
accurate description of the effects of disorder on $T_c$ the critical
value of disorder above which superconductivity is destroyed can also be
obtained by a very simple physical argument.
Let $T_c^{(\text{pure})}(J)$ be the
temperature at which the superconducting transition would occur in the
array of ladders if there were no impurities.
Just above $T_c^{(\text{pure})}$, the thermal length is
$\frac{u_{\rho+}}{T_c^{(\text{pure})}}$ and beyond that length all phase
coherence is lost. Clearly, if the thermal length is smaller
than the localization length in a single
chain containing impurities $\xi_{\text{loc.}}$, phase coherence is lost
before coherent backscattering can build Anderson localization. The
system will escape localization due to the building of the (mean-field)
superconductivity. Thus, if
\begin{equation} \label{simplified}
\frac{u_{\rho+}}{T_c^{(\text{pure})}}< \xi_{\text{loc}}
\end{equation}
Anderson localization will not
suppress the superconducting transition. Equation
(\ref{simplified}) gives a simplified
criterion for the stability of superconductivity.
For fixed Josephson coupling
J, (\ref{MF_general}) leads to $T_c^{(\text{pure})}\sim
J^{\frac{2K_{\rho+}}{2K_{\rho+}-1}}$. Thus, the higher $K_{\rho+}$ the
higher $T_c^{(pure)}$. From the preceding section, the localization
length both in the s-wave and d-wave superconducting phase is
$\xi_{\text{loc.}}\sim (\frac 1 D)^{\frac 2 {3-2K_\rho+}}$. Increasing
$K_{\rho+}$ also reduces $\xi_{\text{loc.}}$. Thus, the two effects
reinforce each other, and make the $SC^s$ phase, that exists for
$K_{\rho+}>1$ more stable against Anderson localization than the
$SC^d$ phase that exists only at $K_{\rho+}<1$.

\section{Experimental consequences}\label{link_w_expts}
The theoretical results obtained in the preceding sections have
important
consequences for experimental systems that are believed to be well
approximated by coupled chains systems, namely the doped ladder
systems which present a superconducting transition and the 2 band
quantum wire. In the former case, one would like to know if
the superconducting transition is related to the divergence of
superconducting fluctuations in the strictly one dimensional system
that results at the mean field level in a finite T superconducting
transition or if the physics of the transition is a two or three
dimensional one. We believe that the resilience of superconductivity
to disorder is a stringent test of effective dimensionality.
In the case of quantum wires, we discuss the experimental consequences
of our results for the conductivity
and charge stiffness in the interacting system. Measurements of the
conductance would allow to check th above theories for the
ladders and provide a measurement of the  Luttinger
liquid parameter in the charge sector, providing some insight on
the strength of interactions in these systems.

\subsection{Superconductivity of doped ladder systems}

Our study has various experimental consequences for the observation of
superconductivity in Ladder systems. First, if the superconductivity is
to come from purely repulsive interactions (i.e. to be of the d-wave
type), it should be extremely sensitive to disorder as we showed in
section~\ref{mfsimple}. In fact, any randomness would induce a
conductivity that never increases as temperature decreases (see figure
\ref{sigma_vs_T}), so that superconductivity  would
be impossible to probe except in extremely pure samples.
Such sensitivity with respect to disorder is certainly
consistent with the difficulty in observing any type of
superconductivity in the ladder systems
$\rm Sr_{n-1}Cu_{n+1}O_{2n}$ \cite{takano_srcuo,takano_spingap}.
However superconductivity seems indeed to be observed in
$\rm Sr_{0.4}Ca_{13.6}Cu_{24}O_{41.36}$ \cite{uchara_srcacuo}
under pressure ($\sim 3 \rm GPa$). Whether such superconductivity
is of the d-wave type is
of course still open. Various experimental facts, however seem to
indicate that if it is the case,
it is unlikely that such superconducting phase could be described
by weakly coupled ladder systems. Indeed one could use the criterion
(\ref{simplified}) to estimate the localization length.
Taking a reasonable
value of $10^6 m s^{-1}$ for the Fermi velocity, one obtains from the
observed $T_c \sim 10K$, a minimal localization length of $\xi \sim
10000 \AA$. Using (\ref{locle2ch}), this leads to extremely long
mean-free paths ($l=\xi$ for $K=1/2$) when one is in the d-wave sector.
So unless the chains are extremely pure, a fact
not likely to be true in such doped materials, one expects based on one
dimensional physics alone that the superconductivity should be totally
suppressed. If the presence of superconductivity is due to an extremely
pure system (which is doubtful) then,
introducing more disorder in the system (for instance by
irradiation) should induce a dramatic decrease of the critical
temperature.

Besides the extreme sensitivity of $T_c$ to disorder other arguments are
again a simple stabilization of one-dimensional physics in the
experimental compound: even if one could be below the
critical disorder
strength determined by (\ref{simplified}) and Figure~\ref{TcD_dwave},
the physics above $T_c$ should be dominated by the one-dimensional
(ladder) effects. In this regime the resistivity goes {\bf up} with
decreasing temperature as described in section~\ref{transport}.
The observed resistivity showing a monotonic {\bf decrease} of the
resistivity (roughly with a $T^2$ or $T$ law)
is again incompatible with the
one-dimensional description. If one is in the purely repulsive sector,
the most likely explanation of the main experimental features
is that under pressure the interchain hopping between the ladders become strong
enough so that the system does not retain its one-dimensional feature,
but is more accurately described by two-dimensional physics.
Such an interpretation is also compatible with the fact that the system
at ambient pressure is insulating.
In that case, the coupled bichains treatment
becomes extremely questionable, and it is probably better to start from
a two-dimensional description, for which disorder effects are probably
weaker, but for which the nature of the superconducting phase has yet to
be completely elucidated.

Another interesting, but probably more farfetched, possibility could be
that the system is in fact in the orbital antiferromagnetic sector.
In such a sector the effects of
disorder are much more reduced, and even very large localization lengths
can lead to reasonable mean-free paths ($\xi/a$ is at worst $(l/a)^2$,
and diverges for $K=3/2$, see (\ref{locoaf})). The resistivity decreases
with temperature according to (\ref{condoaf}). Here the difficulty lies
more in getting the interactions corresponding to this phase, since one
needs local repulsion and a sizeable nearest neighbor attraction. In any
case, careful measurements of the temperature dependence of the
resistivity above $T_c$ could help to decide if such OAF effects are
present. Of course here again, one cannot exclude that the physics is
two-dimensional to start with, but at least now the one-dimensional
starting point is more consistent with the dominant experimental
features.

\subsection{Application to quantum wires}

Progress in nanostructure technologies have allowed for measurements
of the transport properties of low dimensional electronic systems.
In  particular, in recent experiments on quantum wires
\cite{meirav_wires,meirav_gaas_1d,goni_gas1d,tarucha_wire_1d}, the
\emph{conductance} of a quasi one dimensional electron gas has been
measured at very low temperatures. For the pure system, or extremely
weak disorder one finds quantized values of the conductance
\onlinecite{tarucha_wire_1d} in good
agreement with the theoretical predictions\cite{safi_pure_wire,maslov_pure_wire}
at fractions of $\frac h {e^2} $ as a
function of width of the quantum wire (i.e. of the number of subbands
at the Fermi level). The relation between the number of channels
and resistance has been verified \cite{tarucha_wire_1d}.
Impurities on the other hand induce backward scattering that is known
to cause Anderson Localization in a sufficiently long system. In small
enough system, it leads to reduction of conductance as the length of
the system is increased or the temperature is lowered.
Deviations of conductance from $\frac{e^2} h$ as a function of temperature have
indeed been obtained
in experiments\cite{tarucha_quant_cond} as well as deviations as a
function of of the length of the wire \cite{tarucha_wire_1d} and can be
related to the Luttinger liquid exponent. The correction
to the conductance due to impurities
\cite{kane_qwires_tunnel,kane_qwires_tunnel_lettre,ogata_wires_2kf}
is of the form
\begin{equation}\label{k&f_formula}
G(T)=\frac{e^2}h -gT^{-\nu}
\end{equation}
where $\nu=1-K_{\rho}$ is the conductivity exponent
\cite{giamarchi_moriond}.
The derivation is similar to the derivation of the T dependence of conductivity
in appendix \ref{condumem}. For finite size systems $T$ can be replaced
by the lower cutoff $v_F/L$ in (\ref{k&f_formula}).
This formula only holds at high enough
temperatures or for systems of length $L$ shorter than the localization
length for which the corrections term is small.

If two channels are present in the wire, the system becomes then
equivalent to a ladder system. Two band present at the Fermi level are
the equivalent of the bonding and antibonding bands of the ladder
system. On then expect that the whole physics derived in
section~\ref{transport} should apply to these wires. In particular,
since one expects reasonably repulsive interactions one should be in the
$SC^d$ of the $CDW^{4k_F}$ phase. Going from a single chain to the
ladder should have observable consequences on the transport properties .
First since the localization length increases drastically in the ladder
system one would expect the conductance corrections due to disorder, to
be much weaker for two channels. This of course assumes that the
typical interactions do not vary too much when going from one channel to
two channel, a fact which is not certain. Secondly by doing the
expansion of the corrections to
conductance for the ladder and using the conductivity exponent
(\ref{transport}), one would obtain $\nu=2-2K_{\rho+}$.
A fit of the temperature dependence of the conductance in \cite{tarucha_wire_1d},
could allow to extract the Luttinger liquid exponents for the ladder
system, as well as check the above predictions\cite{maslov_qwires_nchannels}.

\section{Conclusion}  \label{concl}

In this paper we have examined the effects of disorder on a 2-legs
ladder system, using RG techniques. We have computed the effects of
disorder on the phase diagram as well as the localization length.
Disorder has drastic effects on the phase diagram. For
spinless fermions, it leads to an extremely strong localization of the
charge density wave phase that exists for repulsive interactions. Such
localization is even stronger than for a single chain. On the other hand
for the ladder system there is a remarkable stability of the s-wave
superconducting phase (for attractive interactions), compared to the
single chain case. The
insulator-superconductor transition occurs in the vicinity of the non
interacting point for a pure t-V model whereas in the one chain system
it occurs for strongly attractive interactions.

For fermions with spin, the repulsive part of the phase diagram is also
strongly localized
by disorder. In particular the novel d-wave superconducting phase found
for ladder systems is completely suppressed by an arbitrarily small
amount of disorder. We emphasize that this is not \emph{only} a pair
breaking effect but a much stronger Anderson localization effect.
On the other hand the s-wave superconducting phase occuring for
attractive
interactions is again much more stable to disorder than its one chain
counterpart.

Besides obtaining the phase diagram, we have also investigated the
transport properties of the t-V and Hubbard two chain systems. The RG
enabled us to compute the localization length and the charge stiffness
as a function of disorder (see tables \ref{table2},\ref{table3}) and
the temperature and frequency dependence of the conductivity. Various
remarkable fact emerged. First the behaviors of the spinless ladder and
the ladder with spins are very different. In particular the
spinless ladder shows
the same tendency than the single spinless chain, namely that attractive
interactions decrease
localization whereas repulsive interactions enhance it. In
the two chain case, that effect is even stronger. For attractive
interactions there is \emph{no localization}, whereas for repulsive
ones the system is much more localized than its one chain counterpart
(the \emph{exponent} in the dependence of localization length with
disorder is changed). On the other hand for the Hubbard ladder
there is no such effect: up to a
prefactor, the localization length with attractive interaction is the
same as for repulsive ones. As a consequence, the corrections to
conductivity stiffness are the same for attractive and for repulsive
interactions.

The temperature dependence of the conductivity follows a power law of
the form $\sigma(T)\sim T^{2-2K_{\rho+}}$, for temperatures above the
localization scale $T_{\text{loc}}$, where $K_\rho$. For the
repulsive side, where for the pure system one would have the d-wave
superconducting phase, $K_{\rho+}<1$, and thus the conductivity
\emph{decreases} as a function of the temperature even well above
$T_{\text{loc}}$. The transport thus shows \emph{no remnant} of the
superconducting behavior one could have naively expected when looking
at the pure system. This remarkable fact illustrates that transport is
in fact controlled by the density fluctuations of the system and
\emph{not} by the existence of not of slowly decreasing superconducting
correlation functions. Ladder system provides evidence of a phase that
is genuinely a d-wave superconductor as far as phase diagram is
concerned, but would from the transport point of view be closer to an
insulator. Of course such an interesting behavior would clearly
deserve more studies. In particular it would be interesting to know how
the correlation between the density fluctuations and the superconducting
one evolves as the number of chain is increased, and how the crossover
to the three dimensional situation occurs. Such a study goes of course
far beyond the goals of the present paper.

We have applied our results to two types of experimental systems.
First our
results should be relevant for quantum wires with two channels. Here the
prediction for the exponent in the conductivity can be directly checked
by the measuring temperature dependence of the conductance of the
system. Note that the conductivity/conductance exponent $2-2K_{\rho+}$
for the ladder systems is different from  the one for
a single channel (or a single chain) $1-K$. Due to the increase of the
localization length when going from one channel to two channels one
would also expect overall smaller corrections to the conductance for a
given strength of the disorder, and roughly constant interactions.
Investigation of systems with more than two chains would be useful in
order to get a better understanding of the role of internal symmetries
and gaps on the transport properties of quasi one dimensional
systems. This is of course also useful in connection with experiments
on quantum wires. In particular, we expect that the behavior of
systems with an even number of legs is dominated by gap formation
whereas the behavior of systems with an odd number of legs should be
closer to the one of a one chain system.

The other experimental system on which our results could be applied are
of course the  coupled bichains where superconductivity
has recently been obtained \cite{uchara_srcacuo} under pressure.
To compare with this system of coupled ladders, it was necessary to
treat the coupling between different ladders, which we did using a
mean-field approximation. The results indicate that even in the presence
of coupling between ladders the d-wave phase is still much too sensitive
to disorder, to be the one experimentally observed. In addition the
observed temperature dependence of the conductivity would be
incompatible
with the one computed here should these systems be dominated by
one-dimensional (ladder) physics. These observations and the fact that
the conductivity occurs under very large pressure tends to indicate
that the mechanism for
superconductivity in these systems is very likely to be of a two or
three dimensional nature and not just the mere stabilization of the
ladder superconducting phase. On the other hand the system without
pressure has a resistivity that could be more compatible with the
localization effects described here. Of course one interesting question
would be whether one can get a one (ladder) to higher dimensional
crossover as the pressure is applied. This of course could only be
decided by a more quantitative comparison with experiments as well as
further theoretical and experimental work. Adding additional impurities,
for example by
irradiation, could allow to distinguish if the system is in a
one-dimensional regime, since one expect much more drastic localization
effects in that case.

Finally a ladder system with spins exhibits an extremely interesting
orbital antiferromagnetic phase. Although such a phase cannot occur in a
pure Hubbard system it can in principle be stabilized if some nearest
neighbor attraction is added. Although such a phase has \emph{no}
superconducting order parameter, it  has
perfect conductivity in the presence of a random potential. Moreover
that perfect conductivity is also robust in the presence of random
hopping both along the chains and perpendicular to the chains. As far as
transport is concerned  this
phase is therefore a one dimensional ``superconductor''. Nevertheless,
it has only subdominant (in the spinless case) or exponentially decaying
(in the case of fermions with spin) superconducting correlations, again
an illustration that looking at the superconducting fluctuations is
\emph{not} a good criterion to determine the transport properties.
Due to the peculiar nature of this phase it would be interesting to check
whether it survives in ladder systems with more than two legs. More
generally it also deserves
further investigation in dimensions higher than one, both in relation of
flux phases  of two dimensional systems and other orbital phases
proposed for the normal state of cuprate superconductors
\cite{varma_review}.

The study of the disorder effects could also be extended in various
directions. In particular a more detailed description of the physics
inside the localized phase would be suitable. However such a description
is beyond the reach of the simple RG calculation. Going to strong
but diluted disorder is also a challenging problem. In particular
understanding the crossover from the results of
our paper to the limit where disorder suppresses gaps altogether in
the system remains yet to be done.

\acknowledgments

We are grateful to L. Balents, A.J. Millis, H.J. Schulz, C.M. Varma for
useful discussions, and H. J. Schulz for making available
Ref.~\onlinecite{schulz_unpublished} to us.
\appendix

\section{Bosonization technique} \label{bosonisation}

In this section, we will give a short review of the bosonization technique
in order to fix the notations. We give the expressions for a single
chain of spinless fermions. For more species of fermions, one can
bosonize each specie individually, and the corresponding expressions are
given in the text.
\subsection{representation of fermion operators in terms of boson ones}
Non-interacting one dimensional spinless fermions on a lattice
are described by the kinetic energy
\begin{equation}
\label{basic_hamiltonian}
H=-t \sum_{n=1}^N (c^\dagger_{n+1}c_n + c^\dagger_n c_{n+1}) =
\sum_{k} \epsilon(k)c^\dagger_{k}c_{k}
\end{equation}
where $\epsilon(k)=-2t \cos k$ and
$c_n=\frac{1}{\sqrt{N}}\sum_{k}e^{\imath k r_{n}} c_k$.

To obtain the asymptotic (low energy, long wavelength) properties of the
system one can linearize the spectrum near the Fermi
``surface'' ($\pm k_F$) and take the continuum limit by introducing
$\psi(x)=\frac{c_n}{\sqrt{a}}$ with $a$ the lattice spacing and $x=na$.
With our definition the $\psi$'s have the commutation
relations of continuum fermion operators.
We define the R (resp. L) (right and left movers) fermions as fermions
with momentum close to $+k_F$  (resp. $-k_F$) as
\begin{equation}
\psi_{R}(x)=\frac{1}{\sqrt{Na}}\sum_{\mid k\mid<\Lambda} e^{\imath kx}
c_{k_F +k}
\end{equation}
and similarly for $\psi_L(x)$ with $k_F \to -k_F$.
$\Lambda$ is a cut-off needed not to double count fermion states,
and imposed by the linearization of the dispersion relation.
All asymptotic properties can be expressed in term of $\psi_{R,L}$. In
particular the full fermion operator becomes $\psi(x)=e^{\imath
k_Fx}\psi_R(x) + e^{-\imath k_F x}\psi_L(x)$. The Hamiltonian
(\ref{basic_hamiltonian}) becomes
\begin{equation} \label{lin_spinless}
H=-iv_{F}(\psi_{R}^{\dagger}\partial_{x}
\psi_{R}-\psi_{L}^{\dagger}\partial_{x}\psi_{L})
\end{equation}
with $v_F=2ta\sin(k_Fa)$.

Due to the separation into two branch of fermions and the
linearization of the spectrum, the Fourier components of the fermion
density operators
\begin{equation}
\label{density_oprttrs}
\rho_{R,L}(q)=\sum_{k}\psi_{R,L,k+q}^{\dagger}\psi_{R,k}
\end{equation}
have boson commutation relations
\cite{haldane_bosonisation,emery_revue_1d,solyom_revue_1d}
\begin{eqnarray}
\label{commutators}
\left[ \rho_{R}(q), \rho_{R}(-q') \right]=-\frac{L}{2 \pi} q
\delta_{q,q'} \nonumber\\
\left[ \rho_{L}(q) , \rho_{L}(-q') \right]=\frac{L}{2 \pi} q
\delta_{q,q'}\nonumber \\
\left[ \rho_{L}(q) , \rho_{R}(-q') \right]=0
\end{eqnarray}

This allows to rewrite (\ref{lin_spinless}) as
\begin{equation}
\label{rho_direct}
H=\pi v_{F}\int dx\left[ \rho_{R}(x)^2 +\rho_{L}(x)^2 \right]
\end{equation}
with $\rho_{s}(x)=\psi_{s}^{\dagger}(x)\psi_{s}(x)$ for $s=L,R$.
Instead of using the density operators themselves it is more convenient
to introduce
\begin{eqnarray}
\label{densites}
\Pi(x)=\rho_{R}-\rho_{L} \nonumber\\
\frac{-1}{\pi}\partial_{x}\phi=(\rho_{R}+\rho_{L})
\end{eqnarray}
Physically, $\Pi$ is a momentum density while $\partial_{x}\phi$ is
proportional
 to the deviation of the fermion density from its average value.
The commutation relations for the $\rho$'s imply that $[\phi(x),\Pi(x')]=\imath
\delta(x-x')$.
Also, the Hamiltonian rewritten in terms of $\Pi$ and $\phi$ is~:
\begin{equation}
\label{free_hamiltonian}
H=\int dx \frac{v_F}{2\pi} \left[(\pi\Pi)^2 +(\partial_{x}\phi)^2\right]
\end{equation}
which is just the continuum limit of the Hamiltonian of a 1D harmonic chain.
Note that the following procedure could have been applied to a more
complicated lattice Hamiltonian than (\ref{basic_hamiltonian}). All
that is needed is that the Fermi surface reduces to two points.
The effectiveness of bosonization stems from the fact that it is possible
to express the fermions operators in terms of  $\Pi(x)$ and $\phi(x)$.
If one introduces $\theta(x)=\pi\int_{-\infty}^{x}\Pi(x')dx'$
one has the following relations:
\begin{eqnarray}
\label{fundamental_relationship}
\psi_{R}(x)=\frac{1}{\sqrt{2\pi\alpha}}e^{\imath[\theta(x)-\phi(x)]}U_{R}\nonumber\\
\psi_{L}(x)=\frac{1}{\sqrt{2\pi\alpha}}e^{\imath[\theta(x)+\phi(x)]}U_{L}
\end{eqnarray}
$\alpha$ being a cutoff the presence of which is imposed by the
cutoff needed in the linearization of the dispersion relations.

the $U_{R}$ and $U_{L}$ are anticommuting operators introduced by Haldane that
annihilate
one fermion at the Fermi level. These operators also anticommute with their
hermitian conjugates.
It can be verified explicitly that those relations reproduce correctly
the commutators of fermion operators. These $U$ operators give in
general corrections vanishing in the thermodynamic limit and can be
safely dropped. On the other hand,
if there are different species of fermions
(such as up and down spin fermions or band degeneracies), one must
bosonize separately each fermion specie using the formulas for spinless fermions.
It is needed to introduce $U_{L,n},U_{R,n}$ operators and their complex
conjugates (n indexing the internal degrees of freedom such as spin)
to enforce proper fermions anticommutation relations.
In order to make that bookkeeping
less tedious\cite{schulz_moriond}, one can introduce $\eta$ operators
such that~:
\begin{eqnarray}
\label {etas}
\eta_{\alpha}\eta_{\beta}+\eta_{\beta}\eta_{\alpha}=
2\delta_{\alpha,\beta}\nonumber\\
\eta^{\dagger}_{\alpha}=\eta_{\alpha}
\end{eqnarray}
where $\alpha=(L,n)$ or $\alpha=(R,n)$
these operators can therefore introduce  minus signs in the various
bosonized expressions.

\subsection{handling the interactions with bosonization}
Let us consider spinless fermions.
Interactions can then be handled straightforwardly:
If one adds a density density coupling of the form $\int dx U \rho(x)^2$
The density can be decomposed in a slowly varying part
$\rho_{R}(x)+\rho_{L}(x)$
and a $2k_F$ part $e^{2\imath k_{F}x}\psi_{L}^{\dagger}(x)\psi_{R}(x)+
\text{H. c.}$. \\
In the Hamiltonian, one retains only the slowly varying terms (the other term
give a zero value when integrated over $x$).
The $2k_F$ always disappear, while the $4 k_F$ can persist in a half
filled lattice
system \cite{nijs_equivalence}.
As a consequence, at a non commensurate filling, the Hamiltonian reduces to:
\begin{equation}
\label{interacting_hamiltonian}
H=\int \frac{dx}{2\pi} \left[uK(\pi\Pi)^2
+\frac{u}{K}(\partial_{x}\phi)^2\right]
\end{equation}
with $uK=v_F$ as a consequence of galilean invariance.
if one makes the rescaling $\phi \to \phi/\sqrt{K}$ and $\Pi \to \Pi \sqrt{K}$
one has the same Hamiltonian as in (\ref{free_hamiltonian}) with the
correct commutation relation for $\phi$ and $\Pi$.
If one computes the physical correlation functions at 0K such as the
$2k_F$ part of the
fermion Green's function
$G(x-x',t-t')=-\imath\langle T\psi_{R}(x,t)\psi^{\dagger}_{L}(x',t')\rangle$
 it is easily seen that $K$ controls their power law decay while $u$
controls the propagation of excitations.
$u$ and $K$ are also related to physical quantities such as the charge
stiffness
\cite{kohn_stiffness} and the compressibility.
More specifically, defining the compressibility by
 $\chi=-\frac1L (\frac{\partial P}{\partial L})_{T}$ ,
 $ P=-(\frac{\partial F}{\partial L})_{T}$ and taking $T\to 0K$,we have~:
\begin{equation}
\label{compressibility}
\chi=\frac{\pi K}{u k_F^2}
\end{equation}
The charge stiffness is defined by ${\mathcal D} =\frac{L}{2}
(\frac{d^2 E(\varphi)}{d^2 \varphi^2})_{\varphi=0}$, $\varphi$ being
 a flux threading the system.
>From that definition, one obtains~:
\begin{equation}
\label{stiffness_bos}
{\mathcal D}=uK
\end{equation}

The case of fermions with internal degrees of freedom is usually more
complicated,
 because some of the interactions cannot be reduced to $(\partial_x \phi)^2$
terms, the most well known example being the backscattering of two
fermions with
opposite spins \cite{solyom_revue_1d,emery_revue_1d}.
Usually, one finds sine-Gordon Hamiltonians of the form:
\begin{equation}
\label{basic_SG}
H_{SG}=\int \frac{dx}{2\pi}
\left[uK(\pi\Pi)^2+\frac{u}{K}(\partial_x\phi)^{2}\right]
+\Delta \int dx \cos(\beta \phi)
\end{equation}
These Hamiltonians can be studied using renormalization group (RG) techniques
 \cite{emery_revue_1d,giamarchi_logs}.
The flow equations for $K$ and $\Delta$ are of the Kosterlitz-Thouless form
\cite{kosterlitz_renormalisation_xy,jose_planar_2d}.
$\Delta$ has scaling dimension $2-\beta^2 K/4$. therefore  a small $\Delta$
 is relevant for $K<8/\beta^2$.
>From the RG equation for $\Delta$ one sees that there are two regimes:
one small $K$ or large enough $\Delta$ regime, where $\Delta$ is relevant
and a large $K$, small enough $\Delta$ regime where $\Delta$ is irrelevant.
When $\Delta$ is irrelevant, the correlation functions keep their
 power law character up to logarithmic corrections \cite{giamarchi_logs}.
on the other hand, if $\Delta$ is relevant, $\phi$ will acquire a non-zero
 expectation value that minimizes the ground state energy and a gap
will be generated.
It can then be shown\cite{emery_revue_1d} that there
 $\langle f(\phi)\rangle \sim f(\langle\phi\rangle)$
and that
$\langle T_{\tau}e^{\imath \alpha \theta(x,\tau)}e^{-\imath \alpha
\theta(0,0)}\rangle
 \sim \exp(-\frac{\sqrt{x^2+(u\tau)^2}}{\xi})$ where $\xi$ is a
correlation length.
These results are used extensively in the paper.

\section{Memory function calculation of ac and dc
conductivity}\label{condumem}

For the sake of clarity, we will explain the technique on the example of
one chain of spinless fermions (technically this is the simplest case),
and then explain how the calculation can be extended to more complicated cases.
First let us describe the memory function approximation
\cite{gotze_fonction_memoire}.
 The conductivity is given by linear response theory as:
\begin{equation}
\sigma(\omega)=-\imath\frac{\chi(0)-\chi(\omega)}{\omega}
\end{equation}
where $\chi(\omega)$ is the current-current response function
 \cite{gotze_fonction_memoire,giamarchi_umklapp_1d}.
The memory function $M(\omega)$ is defined by:
\begin{equation}
\sigma(\omega)=\frac{-\imath \chi(0)}{\omega+M(\omega)}
\end{equation}
This gives the exact formula \cite{gotze_fonction_memoire}~:
\begin{equation}
M(\omega)=\frac{\omega\chi(\omega)}{\chi(0)-\chi(\omega)}
\end{equation}
an expansion\cite{gotze_fonction_memoire} at high frequency and
 small impurity concentration gives~:
\begin{equation}
M(\omega)=\frac{(\ll F;F \gg_{\omega} - \ll F;F
\gg_{\omega=0})/\omega}{-\chi(0)}
\end{equation}
Where $\ll; \gg$ is a retarded correlator evaluated for the pure system
 and $F=[J,H]$ J being the total current.
To use that formalism in the framework of bosonization, we first need an
 expression for the current \cite{giamarchi_umklapp_1d}.
This can be obtained from the definition of the fermion density
$\rho(x)=\rho_0-\frac{\partial_x \phi}{\pi}$ and the current conservation
equation~: $\partial_t \rho+\partial_x j=0$.
One obtains $j(x)=\frac{\partial_t \phi}{\pi}$.
Using the Heisenberg equation of motion for $\phi$ and noting that the total current
 $J=\int dx j(x)$ on finds $J=uK \int \Pi(x)dx$.
The coupling to disorder being:
\begin{equation}
H_{\text{imp}}=\int dx \frac{\xi(x)}{2\pi\alpha}e^{\imath 2\phi(x)} +
\text{H. c.}
\end{equation}
We get $F \propto \int \frac{\xi(x)}{2\pi\alpha}e^{\imath 2\phi(x)} -
\frac{\xi^*(x)}{2\pi\alpha}e^{-\imath 2\phi(x)} $.

This gives~:
\begin{equation}
\overline{\langle T_{\tau}F(\tau)F(0)\rangle}\propto  \int dx D\delta(x)
 (\frac{1}{x^2+(u\tau)^2})^K \propto \tau^{-2K}
\end{equation}

Therefore, $\ll F;F \gg_{\omega} \propto \int d\tau
e^{\imath\omega\tau}\tau^{-2K}
 \propto \omega ^{2K-1}$.
This gives $M(\omega)\propto \omega^{2K-2}$ and for $K>3/2$
$\sigma(\omega) \propto
\omega^{2-2K}$.
The formula we have obtained is valid only at high frequency. We can
get from it a
 high temperature formula by using the dimensional equivalence of
temperature and
frequency ( e. g. $ \hbar \omega \sim k_B T$).
To generalize the calculation to a more complicated case, we must first note
that in the formula for the current, $\phi$ will be replaced by
$\phi_\rho$ in the case
 of two chains of spinless fermions and $\phi_{\rho+}$ in the case of
2 chains of
fermions with spin.
The coupling to disorder being some $\int dx \xi(x) e^{\imath n \phi}
+\text{H. c. }$,
 $n$ depending on the problem at hand, we see that in the general case
we will just
 have to make the replacement $2K \to \frac{n^2}{2}K$ in the formulas giving
 $\sigma(\omega),\sigma(T)$.

\section{Effective random potential in the presence of
gaps}\label{gapgap}

In that section, we will give a derivation of the RG equation for
$D_a$ at $g_f>0$.
We start with the method of \cite{giamarchi_loc}~:
we compute perturbatively the correlation
 function~: $\langle T_\tau e^{\imath\sqrt{2}\phi_\rho(x_1,\tau_1)}
 e^{-\imath\sqrt{2}\phi_\rho(x_2,\tau_2)} \rangle$.
In second order in the random potential, since
$\langle T_\tau \sin(\sqrt{2}\phi_\parallel)(x,\tau)
\sin(\sqrt{2}\phi_\parallel)(0,0)
\rangle \sim e^{-r/l}$, there is no singular contribution.
Therefore, we must go to fourth order. We will drop the combinatorics since
 we are only interested in the renormalization of D.
The fourth order term is of the following form~:
\begin{eqnarray}
D_a^2\int \frac{dx_1 d\tau_1 dx_2 d\tau_2 dx_3 d\tau_3 dx_4
d\tau_4}{(\pi\alpha)^4}
 [\delta(x_1 -x_4)\delta(x_2-x_3)+\delta(x_1-x_2)\delta(x_3-x_4)] \nonumber \\
\langle T_\tau e^{\imath \sqrt{2}[\phi_\rho(x,\tau)+\phi_\rho(x_1,\tau_1)
+\phi_\rho(x_3,\tau_3)-\phi_\rho(x_2,\tau_2)-\phi_\rho(x_4,\tau_4)
-\phi_\rho(0,0)]}\rangle \nonumber \\
 \langle T_\tau
 \sin(\sqrt{2}\phi_\parallel)(x_1,\tau_1)
 \sin(\sqrt{2}\phi_\parallel)(x_2,\tau_2)
 \sin(\sqrt{2}\phi_\parallel)(x_3,\tau_3)
\sin(\sqrt{2}\phi_\parallel)(x_4,\tau_4) \rangle
\end{eqnarray}
The $\phi_\parallel$ will be exponentially small except when $\mid r_1
-r_3 \mid \ll l$
 and $\mid r_2 -r_4 \mid \ll l$ or $\mid r_1 -r_2 \mid \ll l$
and $\mid r_3 -r_4 \mid \ll l$ (the other cases are equivalent
 to these two ones up to a relabeling of dummy integration variables).
It is easily seen that the second case is in fact trivial.
Therefore, the only interesting contribution comes from the first term.
This term reduces to the simple form~:
\begin{equation}
D_a^2 l^2 C \int  {dx_1 d\tau_1 dx_2 d\tau_2} \delta(x_1-x_2)
\langle  T_\tau e^{\imath \sqrt{2}[\phi_\rho(x,\tau)+2\phi_\rho(x_1,\tau_1)
-2\phi_\rho(x_2,\tau_2)
-\phi_\rho(0,0)]}\rangle
\end{equation}
Where C is a constant that depends on the regularization of the
continuum model.
It can be seen that the term that we obtain can be generated by the
following effective
 coupling~:
\begin{equation}
H_{\text{effective}}=\int dx \xi_{\text{eff.}}(x) e^{\imath
\sqrt{8}\phi_\rho} + \text{H. c.}
\end{equation}
with $\overline{\xi_{\text{eff.}}(x)\xi_{\text{eff.}}(x')} =
D\delta(x-x')$ and $D\propto CD_a^2$.
It is clear that for couplings of the form
$e^{\imath\phi_\rho}\cos(\theta_\parallel)$
the same argumentation will be equally valid.
Note that using SCHA approximation gives different results; This is due to the
fact that normal ordering in SCHA is done without taking the presence
of the gaps into
account. Therefore standard scaling, irrespective of the presence of
the gaps always
holds when one uses SCHA.

\section{The Orbital Antiferromagnet in the presence of random
intrachain hopping and random interchain hopping}\label{other_perturbations}
We consider the following two types of random hopping:
a random hopping along the chains:
\begin{equation}\label{random_hopping_intra}
H_{\text{intrachain}}=\sum_{i,\sigma}\left[ \delta t_i^1 (c_{i+1,\sigma,1}^{\dagger}c_{i,\sigma,1}+c_{i,\sigma,1}^{\dagger}c_{i+1,\sigma,1})+\delta t_i^2 (c_{i+1,\sigma,2}^{\dagger}c_{i,\sigma,2}+c_{i,\sigma,2}^{\dagger}c_{i+1,\sigma,2})\right]
\end{equation}
a random interchain hopping \emph{amplitude}:
\begin{equation}\label{random_hopping_inter}
H_{\text{interchain}}=\sum_{i,\sigma} \delta
t_{\perp,i}\left(c_{i,1}^\dagger c_{i,2}+c_{i,2}^\dagger
c_{i,1}\right)
\end{equation}
Where $\delta t_\perp$ is \emph{real}.
Bosonization of (\ref{random_hopping_intra}) leads to an expression
\emph{identical} to the one that obtains by bosonizing a random
on-site potential. It is then evident that the transport properties of
the Orbital Antiferromagnet are the same in the presence of a random
potential or random hopping along the chains.
Bosonization of equation (\ref{random_hopping_inter}) gives the
following expression:
\begin{equation}
H_{\text{interchain}}=\int \frac{2dx}{\pi\alpha} t_\perp^{2k_F}(x) e^{-\imath
\phi_{\rho+}} \left[\imath \sin \phi_{\rho+}\cos \phi_{\sigma-}\cos
\phi_{\sigma+} +\cos\phi_{\rho-}\sin\phi_{\sigma-}\sin\phi_{\sigma+}
\right]+\text{H. c.}
\end{equation}
It is not difficult to see that such term has
exponentially decaying correlations since $\theta_{\rho-}$ develops a
gap. Integration of the massive $\rho-$ mode
leads to a coupling that is identical to the coupling to a random
potential.
Therefore a random amplitude of the hopping term also does not affect
the transport properties of the OAF more severely than a random
potential and thus the ``superconducting'' transport properties of the
OAF are not an artifact of restricting to random potentials. All
physically admissible random perturbations of the 2 chain system lead
to the same limit for localization delocalization ($K_{\rho+}=3/2$)
the same behavior for conductivity as a function of frequency and
temperature and the same dependence of localization length as a
function of disorder.

On the other hand if the random hopping term has a random
\emph{phase}, there is a direct coupling to the OAF order parameter
and then the OAF phase is suppressed. Such terms are allowed for
instance in a tight binding picture only if the phases on the atoms of
the 2 chain system cannot be made real. This could be achieved with a
random magnetic flux in each plaquette of the 2 chain system.

\bibliographystyle{prsty}


\begin{table}
\caption{The 4 sectors of the pure 2 chain spinless fermions model, as
a function
of $K_\rho$ and $g_f$. The average
value of the massive field $\langle\theta_{\parallel}\rangle$
are indicated together with the phase with the most divergent
susceptibility}
\begin{tabular}{ccccc}
 &I&II&III&IV\\
\tableline
$g_f$&+&+&-&- \\
$K_{\rho}$&$<1$&$>1$&$>1$&$<1$ \\
\tableline
$\langle\theta_{\parallel}\rangle=$&$\frac{\pi}{\sqrt{8}}$&
$\frac{\pi}{\sqrt{8}}$&0&0\\
\tableline
phase &$ OAF$ & $SC^s$ & $SC^d$ & $CDW^{\pi}$\\
\end{tabular}
\label{table0}
\end{table}

\begin{table}
\caption{The 4 sectors of the pure 2 chain Hubbard model, as a function
of $K_\rho$ and $g_1$. The average
value of the field developing a gap are indicated together with the
phase with the most divergent susceptibility}
\begin{tabular}{ccccc}
 &I&II&III&IV\\
\tableline
$g_1$&+&+&-&- \\
$K_{\rho+}$&$<1$&$>1$&$>1$&$<1$ \\
\tableline
$\langle\theta_{\rho -}\rangle$&0&0&0&0\\
$\langle\phi_{\sigma +}\rangle$&$\frac{\pi}{2}$&$\frac{\pi}{2}$&0&0\\
$\sigma
-$&$\langle\phi_{\sigma-}\rangle=\frac{\pi}{2}
$&$\langle\theta_{\sigma-}\rangle=0 $&$\langle\phi_{\sigma-}\rangle=0
$&$\langle\theta_{\sigma-}\rangle=\frac{\pi}{2} $\\

phase &$ SC^d$ & OAF & $SC^s$ & $CDW^{\pi}$\\
\end{tabular}
\label{table1}
\end{table}

\begin{table}
\caption{
The conductivities and localization lengths in the  spinless fermions
case. The phases are the ones of the non disordered system that are
turned into localized ones upon introduction of a small disorder.}
\begin{tabular}{ccc}

phase & $L_{\text{loc.}}$ & $\sigma (T)$ \\
\tableline
$PCDW^{4k_F}$   & $\left(\frac{1}{D_a}\right)^{\frac 2 {3-4K_{\rho +}}}$ & $ T^{2-4K_{\rho+}}$ \\
$PCDW ^{\pi} $ & $\left(\frac 1 {D_a}\right)^{\frac 1 {3-K_{\rho+}}}$ & $
T^{2-4K_{\rho+}} $\\
\end{tabular}
\label{table2}
\end{table}

\begin{table}
\caption{The conductivities and localization lengths in the  fermions with spin
case. The phases are the ones of the non disordered system that are
turned into localized ones upon introduction of a small disorder.}
\begin{tabular}{ccc}

phase & $L_{\text{loc.}}$ & $\sigma (T)$ \\
\tableline
OAF,$SC^d$,$SC^s$ & $\left( \frac 1 {D_a}\right)^{\frac 2 {3-2K_{\rho+}}}$ &$T^{2-2K_{\rho+}}$ \\
$CDW ^{\pi} $& $\left( \frac 1 {D_a}\right)^{\frac 2 {6-K_{\rho+}}}$ & $T^{2-\frac{K_{\rho+}}{2}}$ \\
\end{tabular}
\label{table3}
\end{table}
\newpage

\end{document}